\documentclass[journal]{IEEEtran}
\usepackage{amsmath}
\usepackage{amsfonts}
\usepackage{epsfig}
\usepackage{amssymb}
\usepackage{graphics}
\usepackage{}
\usepackage{float}
\usepackage[tight,footnotesize]{subfigure}
\usepackage{amsmath}
\usepackage{slashbox}
\usepackage[T1]{fontenc}
\usepackage{cite}
\usepackage{mathtools}
\usepackage{array}

\makeatletter
\newcommand{\thickhline}{%
    \noalign {\ifnum 0=`}\fi \hrule height 1.2pt
    \futurelet \reserved@a \@xhline
}
\newcolumntype{"}{@{\hskip\tabcolsep\vrule width 1pt\hskip\tabcolsep}}
\makeatother

\begin{document}

\title{Random Interruptions in Cooperation for Spectrum Sensing in Cognitive Radio Networks}

\author{\authorblockN{Younes~Abdi,~\IEEEmembership{Student Member,~IEEE,} and Tapani~Ristaniemi,~\IEEEmembership{Senior Member,~IEEE}}
\thanks{Y.~Abdi and T.~Ristaniemi are with the Faculty of Information Technology, University of Jyv\"askyl\"a, P.~O.~Box 35, FIN-40014, Jyv\"askyl\"a, Finland, Tel. +358 40 7214 218 \mbox{(e-mail:younes.abdi@jyu.fi, tapani.ristaniemi@jyu.fi)}. 

This work has been financially supported by Jyv\"askyl\"a Doctoral Program in Computing and Mathematical Sciences (COMAS). }\vspace{-0.25 in}
}

\maketitle

\begin{abstract}
In this paper, a new cooperation structure for spectrum sensing in cognitive radio networks is proposed which outperforms  the existing commonly-used ones in terms of energy efficiency. The  efficiency is achieved in the proposed design by introducing random interruptions in the cooperation process between the sensing nodes and the fusion center, along with a compensation process at the fusion center. Regarding the hypothesis testing problem concerned, first, the proposed system behavior is thoroughly analyzed and its associated \emph{likelihood-ratio test} (LRT) is provided. Next, based on a general linear fusion rule, statistics of the global test summary are derived  and  the sensing  quality  is characterized in terms of the probability of false alarm and the probability of detection. Then, optimization of the overall detection performance is formulated according to the \emph{Neyman-Pearson criterion} (NPC) and it is discussed that the optimization required is indeed a decision-making process with uncertainty which incurs prohibitive computational complexity. The NPC is then modified to achieve a good affordable solution by using semidefinite programming (SDP) techniques and it is shown that this new solution is nearly optimal according to the  \emph{deflection criterion}. Finally, effectiveness of the proposed architecture and its associated SDP are demonstrated by simulation results.   
\end{abstract}


\section{Introduction}
\PARstart{E}{fficient} utilization of available resources is a vital requirement in designing modern communication systems. 
In order to increase the efficiency, significant levels of flexibility and adaptability are realized in wireless networks by employing the  cognitive radio (CR) technology which is developed through adding certain artificial-intelligence-based capabilities, such as self-awareness, context-awareness, and machine learning to software-defined radios \cite{mitola2000}. Due to these prominent capabilities, CR has been of great interest  for developing green and spectrum-efficient wireless communication systems, see e.g., \cite{Akyildiz11} and references therein.

Reliable spectrum sensing is of major importance in designing CR networks (CRNs). The average throughput of the secondary users (SUs), their consumed energy, and the amount of interference experienced by the primary users (PUs) are directly related to the effectiveness of the sensing methods incorporated in CRs. In fact, a reliable sensing algorithm is necessary to obtain suitable transmission opportunities without compromising the integrity of the primary network, particularly in low signal-to-noise ratio (SNR) regimes \cite{Paysarvi-Hoseini11}. Nevertheless, impairments such as shadowing and multipath fading associated with typical wireless environments often degrade the spectrum sensing performance. In order to mitigate this issue, the reliability of the sensing process is significantly enhanced through establishing cooperation among spatially-divers sensing nodes. This cooperation, referred to as cooperative spectrum sensing, can be realized in different configurations among which the so-called centralized cooperative sensing is a commonly-used one \cite{Akyildiz11}.  In this structure, the cooperation is coordinated by and the overall sensing outcome is generated in a special node called the fusion center (FC), see e.g., \cite{Abdi14, Chaudhari13, Taricco11, Quan10, Quan08}. This cooperation is in fact performed as a three-phase process, i.e., each node first performs spectrum sensing individually by using its own built-in sensing scheme. Then, the sensing nodes send their local sensing outcomes to the FC through the so-called reporting channels and finally, the FC combines the received local sensing outcomes to decide the presence or absence of the PU. 

Censoring \cite{Rago96, Lunden09, Chen10, Appad08, Maleki13, Maleki13-2, Maleki11, Maleki15} is a common design strategy for reducing the energy consumption of cooperative spectrum sensing schemes. In this method sensors are assumed to censor their observations so that each sensor sends to the FC only informative observations, and leaves those deemed uninformative untransmitted. In hard-decision-based cooperation with censoring \cite{Maleki13, Maleki13-2, Maleki11, Maleki15}, instead of comparing the local sensing outcome with a single threshold to make the binary decisions, it is compared against two thresholds, i.e., an upper threshold and a lower one. If the sensing outcome is below the lower threshold, the sensing node reports the null hypothesis by sending, let's say, 0 to the FC and if the local sensing outcome is above the upper threshold, then the sensing node reports the alternative hypothesis by sending 1 to the FC. However, if the sensing outcome resides between the two thresholds, it is not considered reliable enough to make a local decision and the CR node does not report anything to the FC, saving the energy and communication resources. In soft decision based cooperation \cite{Rago96, Lunden09, Chen10}, each sensing CR node compares the sensing outcome with a predefined threshold and if it is below the threshold, then the sensor avoids reporting to the FC.

Sensor selection \cite{Xia10, Pham10, Najimi13, Monemian14} and sensor scheduling \cite{Xing15, Cardei05, Jiming10, Lee11, Gil11} algorithms provide a different set of methods for improving the energy efficiency in distributed detection schemes. Sensor selection algorithms mainly aim at selecting a set of sensors whose collaboration contributes significantly to the overall detection process. The goal of sensor scheduling algorithms in general is to increase the energy efficiency of cooperative sensing by dividing the sensors into non-disjoint groups which activate successively. In other words, only the sensors from the current active set are responsible for sensing and for reporting the sensing outcome while all other nodes are in a low-energy sleep mode. 

\vspace{-0.3 cm}
\subsection{Related Work}\label{section Related work}
\emph{1) Cooperation in Spectrum Sensing}: 
Modeling and performance optimization of cooperative sensing schemes has long been of great interest. Several cooperation methods have been introduced and analyzed in the literature, such as the scheme based on voting rules \cite{Aalo92} as a simple suboptimal solution, or the so-called OR logic operation \cite{Ghasemi05} and AND logic operation \cite{Visotsky05}. The likelihood-ratio test (LRT) \cite{Poor, Chen05} is known as the optimal fusion method when the distributed nodes report their sensing outcomes to the FC through nonideal analogue communication links \cite{Quan08, Chaudhari12}. The effect of reporting channel impairments on the overall sensing performance has been investigated in \cite{Chaudhari13, Chaudhari12, Chaudhari11, Chaudhari09}. Particularly, the authors in \cite{Chaudhari12} compare the performances of the hard-decision- and soft-decision-based fusion methods and illustrate that, in general, the soft decision significantly outperforms the hard decision when nonideal reporting channels are considered.  Although LRT is commonly considered as the best soft-decision-based cooperation, finding the optimal LRT thresholds for individual nodes and for the FC incurs prohibitive computational complexity \cite{Quan08, Chaudhari12}. Consequently, optimal linear combining \cite{Taricco11, Quan10, Quan08} has been suggested as a very good alternative which provides nearly-optimal results but at affordable computational cost. As a fast alternative linear combining approach with a very good performance, the so-called \emph{deflection criterion} \cite{Picinbono95, Poor}, a.k.a., \emph{deflection coefficient} (DC)  \cite{Derakhshani11} or its modified version, \emph{modified deflection coefficient} (MDC) \cite{Quan09, Quan08, AbdiPIMRC14, Abdi14} is commonly used in the literature to design linear fusion schemes in cooperative sensing scenarios. 
Moreover, the definition of DC is extended in \cite{Abdi14} and  \cite{AbdiPIMRC14}  to enable joint optimization of reporting and fusion processes in cooperative spectrum sensing.

\emph{2) Censoring in Cooperative Sensing}: Censoring-based distributed detection is  investigated in \cite{Appad08} where the system model is provided both in the Bayesian and Neyman-Pearson settings, a so-called censored likelihood ratio test is defined, and  censoring regions are determined for a certain overall detection performance.  In \cite{Appad05}, performance of OR-based cooperative sensing with censoring is optimized  subject to an upper bound on the cost of local sensing and reporting processes. The results provided in \cite{Rago96} show that the communication resources required for reporting is significantly reduced when a censoring-based cooperation is established between the sensing nodes. Efficient censoring-based cooperative sensing under bandwidth constraint is discussed in \cite{Chunhua07} where performance of hard-decision based spectrum sensing is studied with both perfect and imperfect reporting channels. In \cite{Chen10}, analytical expressions for the sensing parameters are derived according to a Neyman-Pearson setup for censoring with both soft and hard fusion schemes, but no constraint on the energy consumption is taken into account. In \cite{Maleki13} censoring in a wireless sensor network is considered where all sensors experience the same SNR and have the same performance while TDMA is assumed for reporting the local decisions to the FC through ideal, i.e., error-free channels. Based on this model, the throughput is maximized for a k-out-of-N fusion rule under a constraint on energy consumption. The censoring effect in hard-decision-based cooperative sensing is investigated in \cite{Maleki13-2} with two different sensing methods. In the first method, the number of samples used by each sensor is fixed whereas in the second approach a so-called sequential sensing technique is used where the sensors sequentially collect observation samples until they reach a decision about the presence or absence of the PU. The latter approach is shown to be more energy efficient than the former one for a given performance. A combined sleeping and censoring method is proposed in \cite{Maleki11}  which serves as an energy efficient technique for cognitive sensor networks with hard decision fusion. The objective in \cite{Maleki11} is to minimize the energy consumed in distributed sensing with OR fusion rule subject to constraints on the detection performance, by optimally choosing the sleeping and censoring design parameters. The constraint on the detection performance is given by a minimum target probability of detection and a maximum permissible probability of false alarm while the local SNRs at the sensing nodes are assumed to be the same for analytical tractability. The combined censoring and sleeping approach is further investigated in \cite{Maleki15} where sensing parameters are derived by minimizing the maximum average energy consumption per sensor subject to a lower-bound on the overall probability of detection and an upper-bound on the overall probability of false alarm for both OR and AND fusion rules. Compared with \cite{Maleki11} which intends to minimizes the network energy consumption, in \cite{Maleki15} minimizing the maximum average energy consumption per sensor is considered.

\emph{3) Sensor Selection/Scheduling Algorithms}: In \cite{Xia10, Pham10} sensor selection methods  are developed to increase the energy efficiency of cooperative sensing in CRNs with hard decision fusion. They concentrate on finding one subset that has the minimum energy consumption.  A sensor selection problem for cognitive wireless sensor networks with hard decision fusion rule and error-free reporting channels is considered in \cite{Najimi13} where the focus is on maximizing the network lifetime. In \cite{Monemian14} an energy-based sensor selection algorithm is considered in a cognitive wireless sensor network to provide approximately the same lifetimes for sensors via  appropriate design of cooperative sensing with hard decision fusion. It is assumed in \cite{Monemian14} that all of the sensors experience the same SNR and operate with the same sensing thresholds. An energy efficient sensor scheduling algorithm is proposed in \cite{Xing15} to optimally schedule the activities of  sensors to provide the required sensing performance and increase the overall secondary system throughput. This sensor scheduling method is based on OR fusion rule and assumes error-free reporting channels.  Sensor scheduling schemes in \cite{Cardei05, Jiming10, Lee11, Gil11} aim at extending the network lifetime while maintaining full coverage in a certain area. 
\vspace{-0.3 cm}
\subsection{Contribution}\label{section Contribution}
In this paper, we extend the works in \cite{Taricco11, Quan10, Quan08} and propose a novel energy-efficient cooperative sensing scheme for CRNs. This new design is achieved by adding two new mechanisms to the commonly-used cooperative sensing structure. The first mechanism is realized as a set of controllers mounted on the sensing nodes to make random energy-saving interruptions in their cooperation with the FC. The second mechanism is a compensation process at the FC. This compensation, which is realized as a linear estimator, aims at recovering the local test summaries out of degradations caused jointly by the interruptions and reporting channel contaminations. The estimation of the local test summaries is realized in the proposed system by using the spatio-temporal cross-correlations of the sensor outcomes as well as auto-covariance functions characterizing the behavior of the reporting channels. 

We model and thoroughly analyze the proposed system to derive the global test summary statistics in terms of probability distributions based on which the controllers work. By using the statistics obtained, we formulate  the proposed system performance optimization according to two commonly-used criteria, namely the Neyman-Pearson criterion (NPC) and the deflection criterion. 
Specifically, first, we analyze the global test summary statistics and characterize the system performance by providing closed-form relations for the probability of false alarm and the probability of detection. We then derive the detection threshold for a fixed false alarm probability and formulate the system performance optimization based on NPC and subject to a constraint on the energy consumed at the local-sensing and reporting phases. In consequence, we formulate a significant tradeoff concerning the overall detection quality along with joint energy consumption of the local sensing and reporting phases. We discuss that the obtained optimization is inherently an stochastic program with prohibitive computational complexity and then, we consider the optimization as a decision making process with uncertainty. 
By using this decision making structure, we show the link between the NPC and DC for the proposed cooperative sensing scheme.  

Regarding the DC, we formulate the performance optimization as a convex-over-convex fractional programming problem. This fractional program, which is solved by a branch-and-bound (BnB) algorithm, gives optimal distributions of random interruptions for maximizing the overall detection performance, subject to a constraint on the energy consumed at the local-sensing and reporting phases. Through delicate algebraic manipulations, we convert the fractional program into a set of nonconvex quadratically-constrained quadratic programs (QCQP) whose solution can be obtained in polynomial time by using standard techniques in nonconvex quadratic programming. 
Therefore, we construct an effective computationally-affordable method for solving the proposed optimization even when dealing with large-scale networks.

\vspace{-0.3 cm}
\subsection{Discussion and Comparison}\label{section DiscussCompare}
 In the proposed method the local sensing and reporting processes in a CR node are occasionally suspended if the contribution of that node to the overall sensing process does not significantly improve the system performance. As explained in the following, the result of such an interruption is more than just saving the energy of the sensing nodes.

Spectrum sensing in CRNs is often concerned with protecting several, and in many cases, heterogeneous PUs. Moreover, the sensing process is often performed on multiple frequency bands. When a CR node is not able to significantly contribute to the overall sensing process, this generally means that one or maybe more PUs operating on a certain group of frequency bands are hidden from this particular sensing node. However, this CR node may still be able to detect signals from other PUs operating possibly on a different geographical location and/or on a different frequency band. Therefore, instead of performing spectrum sensing continuously and reporting unreliable results which will most-likely be discarded at the FC, it is more efficient for this CR node not to waste its resources in sensing the signals from the hidden PUs and instead, to allocate its signal processing and reporting resources to discover signals from other PUs which are potentially not hidden.

Therefore, by using the proposed cooperation scheme, the sensing nodes can spend their available resources more effectively and obtain a better cooperation performance based on the available resources. Note also that, this efficiency cannot be achieved by the optimal linear cooperation in which the sensing nodes continuously perform their spectrum sensing and it is up to the FC to decide about their contribution by assigning weights to the received reports.

The censoring-based methods are not as effective as the proposed random interruptions either. This is due to the fact that, in cooperative sensing with censoring, a sensing node continuously performs the local sensing process regardless of the quality of the PU signal at its receiver, and only the reporting process is avoided when the sensing node is not able to reliably detect the PU signal. Therefore, in a censoring-based cooperation the CR nodes with unreliable sensing outcomes are not able to save their local sensing resources or use them to sense other PU signals or other spectrum bands. 

In sensor selection or sensor scheduling algorithms, energy efficiency is achieved by giving binary send/no-send instructions to the sensing nodes. Despite the variety of objective functions used in their performance optimization, these algorithms can be considered as a special case compared to the proposed interruption-based cooperation. The reason is that even though the efficiency is realized in the proposed method based on binary sleep/wakeup
instructions, the overall contribution of each node is not affected in a binary manner. As a simplified explanation (see Section \ref{subsection Model} and Section \ref{section NPC} for details), since we are dealing with the statistics of the received reports which go through a compensator and a soft fusion process, the contribution of each node (let's say node $i$), is affected (i.e., multiplied) by a continuous factor (i.e., the Bernoulli probability $p_i$). Therefore, the reporting power and local computation cost are scaled down by a continuous factor, compared to a simple send/no-send structure which turns the sensing node either on or off. Nevertheless, the proposed method still enjoys the benefits of an on-off power saving scheme due to its unique sleep/wakeup structure.

To the best of our knowledge, the present work is the first one to consider energy efficiency in a CRN by optimizing the detection performance of cooperative spectrum sensing with soft fusion rule at the FC, nonideal reporting channels, and different SNR levels at the sensing nodes which do not censor their local sensing outcomes.

\subsection*{Organization}\label{section Organization}
The rest of this paper is organized as follows. In Section \ref{section System Model and Analysis}, general modeling assumptions and details of the proposed system are introduced and its behavior is analyzed. In Section \ref{section NPC}, the system performance optimization is formulated in a Neyman-Pearson setting and the challenges in the optimization are discussed. In Section \ref{section SO}, a scenario-tree-based stochastic programming method is developed for the problem. The  stochastic programming approach is then linked to the DC in Section \ref{section DC} where a low-complexity effective solution for the problem based on standard semidefinite programming methods is constructed. The effectiveness of the proposed cooperative spectrum sensing is then demonstrated through simulation results in Section \ref{section Numerical Results}, followed by concluding remarks provided in Section \ref{section Conclusion}.

\section{System Model and Analysis}\label{section System Model and Analysis}
In this section, we first introduce our notation along with a set of definitions which facilitate mathematical representation of the proposed analysis. Then, the system model is introduced in detail and the parameters needed for performance optimization are derived accordingly. 
\vspace{-0.3 cm}
\subsection{Notation}\label{subsection Definitions}
Matrices and column-vectors are denoted in boldface by uppercase and lowercase letters, respectively. The notation $\left \| \mathbf{x}  \right \|$ denotes the Euclidean norm of the vector $\mathbf{x}$. The notation $\textup{diag}(\mathbf{x})$ represents a diagonal matrix whose main diagonal is $\mathbf{x}$.  Matrix inequality is represented by $\succeq $, i.e., $\mathbf{A}\succeq\mathbf{B}$ means that $\mathbf{A}-\mathbf{B}$ is positive semidefinite and for a vector it denotes the element-wise inequality with $\succ$ representing the strict inequality. 
The identity matrix is denoted by $\mathbf{I}$ and $\mathbf{0}_n$ denotes the $n \times n$ null matrix while  $\mathbf{1}$ and $\mathbf{0}$ denote all-ones and all-zeros column vectors, respectively. $\mathbf{e}_i$ denotes a column vector where all elements are zero except for the $i$th element which is one. $\textup{Tr}(\mathbf{A})$ refers to the trace of $\mathbf{A}$, i.e., sum of elements in the main diagonal of $\mathbf{A}$. The vectorization of matrix $\mathbf{A}$ is denoted by $\textup{vec}\left ( \mathbf{A} \right ) $ which represents a vector obtained by stacking the columns of $\mathbf{A}$ on top of one another. $\mathbf{A}\circ \mathbf{B}$ represents the Hadamard product of matrices $\mathbf{A}$ and $\mathbf{B}$.

In order to account for temporal and spatial representations of signals, we use the following notation. For vector $\mathbf{x}$ in the following form
\begin{equation}\label{N01} 
\mathbf{x}(m) = [x_1(m), ... , x_K(m)]^T  
\end{equation}
we consider these three notations
\begin{equation}\label{N02} 
\mathbf{X}_L(m)\triangleq \left [ \mathbf{x}(m),\mathbf{x}(m-1), ... , \mathbf{x}(m-L)  \right ]^T 
\end{equation}
\begin{equation}\label{N03} 
\mathbf{x}_L(m) \triangleq \textup{vec}\left ( \mathbf{X}_L(m) \right )  
\end{equation}
\begin{equation}\label{N04} 
\mathbf{\tilde{X}}_L(m)\triangleq \textup{diag}\left ( \mathbf{x}_L(m) \right )
\end{equation}
Note that we represent the time index by $m$. 

For the two hypotheses considered in this paper, i.e., the null hypothesis $\mathcal{H}_0$ corresponding to the absence of the PU signal and $\mathcal{H}_1$ representing the presence of the PU signal, the following notations are used to represent  conditional second-order statistics of signals (for $h = 1,2$)
\begin{align}\label{N05} 
\mathbf{C}_{\mathbf{x}|\mathcal{H}_h}&\triangleq E\left [(\mathbf{x}-\boldsymbol{\mu}_{\mathbf{x}|\mathcal{H}_h})(\mathbf{x}-\boldsymbol{\mu}_{\mathbf{x}|\mathcal{H}_h})^H \right|\mathcal{H}_h ] \nonumber \\
&= \mathbf{R}_{\mathbf{x}|\mathcal{H}_h}-\boldsymbol{\mu}_{\mathbf{x}|\mathcal{H}_h}\boldsymbol{\mu}_{\mathbf{x}|\mathcal{H}_h}^H
\end{align}
\begin{align}\label{N06} 
\mathbf{C}_{\mathbf{xy}|\mathcal{H}_h}&\triangleq E\left [(\mathbf{x}-\boldsymbol{\mu}_{\mathbf{x}|\mathcal{H}_h})(\mathbf{y}-\boldsymbol{\mu}_{\mathbf{y}|\mathcal{H}_h})^H \right|\mathcal{H}_h ] \nonumber \\
&= \mathbf{R}_{\mathbf{xy}|\mathcal{H}_h}-\boldsymbol{\mu}_{\mathbf{x}|\mathcal{H}_h}\boldsymbol{\mu}_{\mathbf{y}|\mathcal{H}_h}^H
\end{align}
where $\boldsymbol{\mu}_{\mathbf{x}|\mathcal{H}_h}$ and $\boldsymbol{\mu}_{\mathbf{y}|\mathcal{H}_h}$ denote mean of $\mathbf{x}$ and mean of $\mathbf{y}$, conditioned on $\mathcal{H}_h$, respectively. The main diagonal of the autocorrelation matrix $\mathbf{R}_{\mathbf{x}|\mathcal{H}_h}$ is denoted  $\boldsymbol{\mathfrak{R}}_{\mathbf{x}|\mathcal{H}_h} $ which is defined as
\begin{equation}\label{N07} 
\left (\boldsymbol{\mathfrak{R}}_{\mathbf{x}|\mathcal{H}_h}  \right )_{i,j}\triangleq \delta_{i,j}\left ( \mathbf{R}_{\mathbf{x}|\mathcal{H}_h} \right )_{i,j}
\end{equation} 
where $\delta_{i,j}$ is Kronecker's delta function which equals to 1 if $i=j$ and 0 otherwise. 

Non-conditional statistics are related to their conditional counterparts according to the following definitions
\begin{equation}\label{N08} 
\mathbf{C}_{\mathbf{x}}\triangleq \textup{Pr}\{\mathcal{H}_0\}\mathbf{C}_{\mathbf{x}|\mathcal{H}_0}+\textup{Pr}\{\mathcal{H}_1\}\mathbf{C}_{\mathbf{x}|\mathcal{H}_1}
\end{equation} 
\begin{equation}\label{N09} 
\mathbf{C}_{\mathbf{xy}}\triangleq \textup{Pr}\{\mathcal{H}_0\}\mathbf{C}_{\mathbf{xy}|\mathcal{H}_0}+\textup{Pr}\{\mathcal{H}_1\}\mathbf{C}_{\mathbf{xy}|\mathcal{H}_1}
\end{equation}
Note that other non-conditional statistics in this paper are related to their conditional counterparts in a similar fashion. 

As another definition regarding the second-order statistics, we have the following $K(L+1)\times K$ matrix
\begin{align}\label{N10} 
\mathbf{C}_{\mathbf{x}_L\mathbf{y}} & \triangleq  E\left [ \mathbf{x}_L(m)y_1^{\dagger}(m), ... , \mathbf{x}_L(m)y_K^{\dagger}(m) \right ] \nonumber\\
&-\left [ \boldsymbol{\mu}_{\mathbf{x}_L}y_1^{\dagger}(m), ... , \boldsymbol{\mu}_{\mathbf{x}_L}y_K^{\dagger}(m)  \right ]
\end{align}
where $\boldsymbol{\mu}_{\mathbf{x}_L} \triangleq E[\mathbf{x}_L] $ and $\dagger$ denotes complex conjugation. 

Scalar conditional autocorrelation and autocovariance functions of $\mathbf{x}(m)$ for $k=1,2,...,K$, $n = 1,2, ..., K$, $l=0, 1, ... , L$, $r=0, 1, ... , L$, and $h=0,1$ are denoted by
\begin{equation}\label{N011} 
\omega_{\mathbf{x}|\mathcal{H}_h}(k,n;l,r)\triangleq E\left [ x_k(m-l) x_n^{\dagger}(m-r)|\mathcal{H}_h\right ]
\end{equation} 
\begin{align} \label{N12} 
c_{\mathbf{x}|\mathcal{H}_h}(k,n;l,r) & \triangleq  \omega_{\mathbf{x}|\mathcal{H}_h}(k,n;l,r)\nonumber\\
&-E\left [ x_k(m-l)|\mathcal{H}_h\right ]E\left [ x_n^{\dagger}(m-r)|\mathcal{H}_h\right ]
\end{align}
Note that with this definition for the autocovariance function we have (for $i = 1, ... , K$ and $j = 1, ..., K$)
\begin{equation} \label{N13} 
\left (\mathbf{C}_{\mathbf{x}|\mathcal{H}_h}  \right )_{i,j} = c_{\mathbf{x}|\mathcal{H}_h}(i,j;0,0)
\end{equation}
However, regarding the covariance matrix of $\mathbf{x}_L$, which is denoted by $\mathbf{C}_{\mathbf{x}_L|\mathcal{H}_h}$, we have (for $i = 1, ... , K(L+1)$ and $j = 1, ..., K(L+1)$)
\begin{equation} \label{N14} 
\left (\mathbf{C}_{\mathbf{x}_L|\mathcal{H}_h}  \right )_{i,j} = c_{\mathbf{x}|\mathcal{H}_h}(k,n;l,r)
\end{equation}
where $k = \left \lceil \frac{i}{L+1}  \right \rceil$, $n = \left \lceil \frac{j}{L+1}  \right \rceil$, $l = (i+L)\textup{mod}(L+1)$, and $r = (j+L)\textup{mod}(L+1)$. $\left \lceil \cdot  \right \rceil$ and $\textup{mod}$ denote the ceiling function and modulo operation, respectively. 

Now we are ready to model and analyze the proposed system behavior. 
\vspace{-0.3 cm}
\subsection{System Model}\label{subsection Model}
\begin{figure}[]
  \hspace{0cm} \vspace{-5 mm}  \includegraphics[scale=0.3]{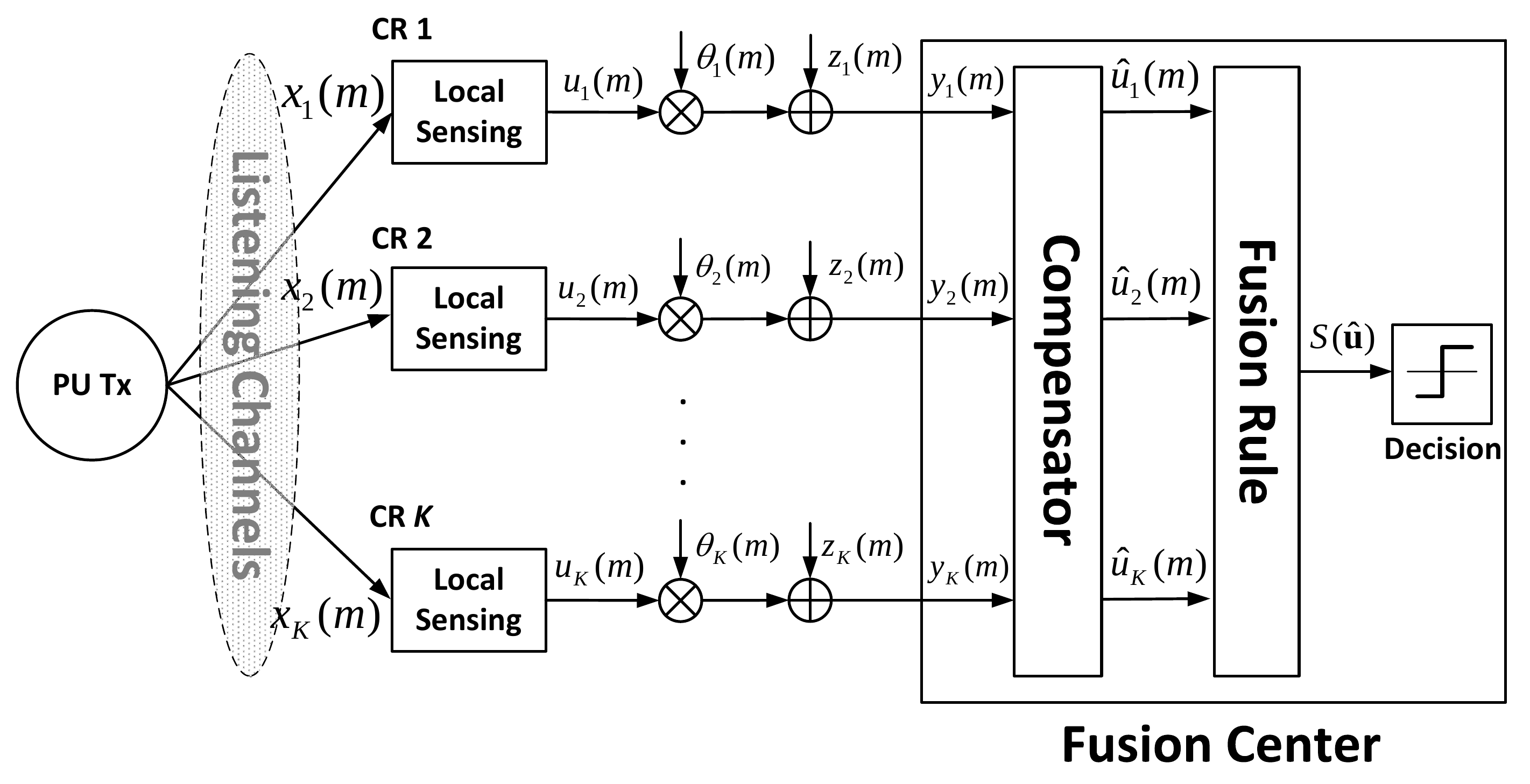}\\
  \caption{The proposed system architecture.}\label{fig1}
\end{figure}

A CRN with $K$ sensing nodes is considered. These nodes cooperatively sense the radio spectrum to find temporal and/or spatial vacant bands for their data communication. Fig. \ref{fig1} shows the major elements of the proposed design. Each CR node is equipped with a built-in spectrum sensor which enables it to detect the PU signal through inspecting its own listening channel. Listening channels are referred  to the channels between the PU and the sensing nodes. The sensing nodes have access to a dedicated but nonideal reporting channel to send their individual sensing outcomes to the FC.

In our adopted model, the $m$th sample of the received PU signal at the $i$th CR node is represented as
\begin{eqnarray}\label{E15} \left\{ 
  \begin{array}{l l}
   x_{i}(m) = \nu_{i}(m), & \quad \mathcal{H}_0\\
   x_{i}(m) = h_{i}s(m) + \nu_{i}(m), & \quad  \mathcal{H}_1
  \end{array} \right.
\end{eqnarray}
where $s(m)$ denotes the signal transmitted by the PU and  $x_{i}(m)$ is the received signal by the $i$th SU. $ h_{i}$ is the listening channel block fading coefficient, which is assumed to be constant during the detection interval. Listening channel gains are assumed to be independent circularly-symmetric Gaussian random variables. $\nu_{i}(m)$ denotes the circularly-symmetric zero-mean additive white Gaussian noise (AWGN) at the CR sensor receiver, i.e., $\nu_{i}(m) \sim \mathcal{CN}(0,\sigma_{\nu_{i}}^{2})$. $s(m)$  and $\left \{ \nu_{i}(m) \right \}$ are assumed to be independent of each other. 

\begin{figure}[]
  \hspace{1.1cm} \vspace{-5 mm} \includegraphics[scale=0.3]{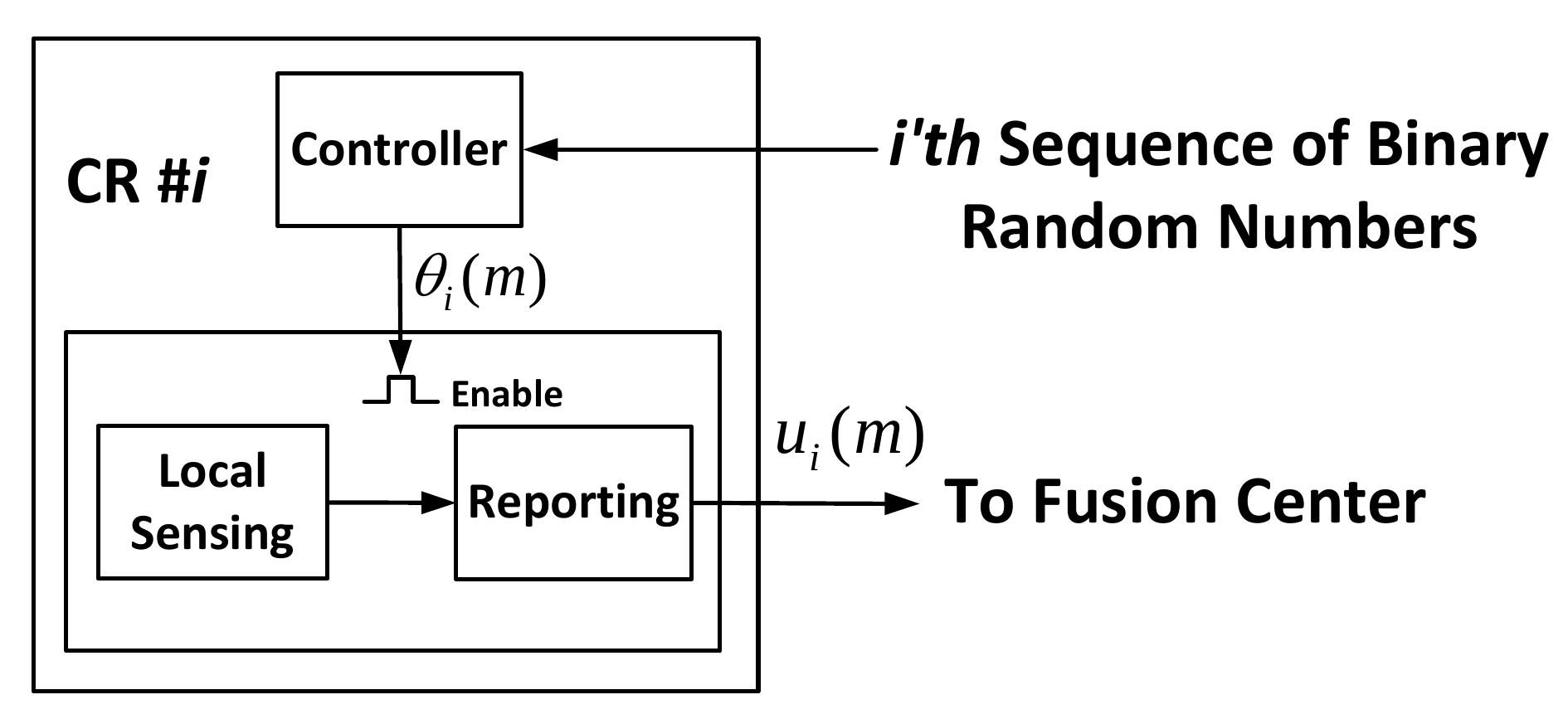}\\
  \caption{Controlling the local sensing and reporting processes at the sensing nodes based on the predetermined sequence of binary random numbers which are generated and shared with CR nodes when initializing the cooperation. } \label{fig1_1} \vspace{-0.3 cm}
\end{figure}

In this cooperative sensing scheme, the behavior of each node is controlled by a predetermined sequence of binary random numbers generated at the FC. Whenever the cooperative sensing is performed, only the nodes whose corresponding random number is one are allowed to contribute to the overall sensing process. Specifically, if the $i$th CR is among the cooperating nodes, $i\in \{1, 2 , ... , K\}$, it  performs spectrum sensing by using its built-in sensor to derive a local test statistic $u_{i}$. Without loss of generality, the local sensing process is assumed to be in the form of energy detection which can be represented as 
\begin{eqnarray}\label{E16} 
u_i(m) = \sum_{k=0}^{N-1}\left | x_i(Nm-k) \right |^2
\end{eqnarray}
The resulting sensing outcome is then transmitted to the FC through the reporting channel. We represent this \emph{interrupted cooperation} as
\begin{equation}\label{E17} 
  y_i(m) = \theta_i(m)u_i(m)+z_i(m)
\end{equation}
where $y_i(m)$ denotes the $i$th received sensing outcome at the FC, $\theta_i(m) \in \{0,1\}$ denotes the random number which controls the $i$th sensing node, and $z_i(m)\sim \mathcal{CN}(0,\sigma_{z}^{2})$ is the reporting channel contamination modeled as AWGN. It is worth noting that, for simplicity, we have not explicitly shown the controllers in Fig. \ref{fig1}. However, (\ref{E17}) clearly demonstrates that when the generated random number $\theta_i$ is zero, nothing is transmitted to the FC. This process can be implemented in practice by using an energy-saving controller which suspends the local sensing and reporting processes when the corresponding random number is zero, see Fig. \ref{fig1_1}.

In the matrix form we can represent (\ref{E17}) as
\begin{equation}\label{E18} 
 \mathbf{y}(m)=\textup{diag}(\boldsymbol{\theta}(m))\mathbf{u}(m)+\mathbf{z}(m)
\end{equation}
where $\mathbf{y}(m) \triangleq \left [ y_1(m), ... , y_K(m) \right ]^T$, $\boldsymbol{\theta}(m) \triangleq \left [ \theta_1(m), ... , \theta_K(m) \right ]^T$, $\mathbf{u}(m) \triangleq \left [u_1(m), ... , u_K(m) \right ]^T$, and $\mathbf{z}(m) \triangleq \left [ z_1(m), ... , z_K(m) \right ]^T$. In the proposed system the random numbers $\theta_i$, $i=1, ... , K$, are generated independently over time and space, i.e., $\theta_k(m-l)$ and $\theta_n(m-r)$ are independent if $k \neq n$ or $l \neq r$. Moreover, $\theta_i(m)$ follows the Bernoulli distribution with $p_i(m) \triangleq \textup{Pr}\left \{ \theta_i(m)=1 \right \}$. Therefore, the autocorrelation and autocovariance functions of $\boldsymbol{\theta}$ can be represented as
\begin{align}\label{E19} 
\omega_{\boldsymbol{\theta}}(k,n;l,r) &= p_k(m-l)p_n(m-r) \nonumber\\
&+ \delta_{k,n}\delta_{l,r}p_k(m-l)[1-p_k(m-l)]
\end{align}
\begin{align}\label{E20} 
c_{\boldsymbol{\theta}}(k,n;l,r) &= \omega_{\boldsymbol{\theta}}(k,n;l,r) -p_k(m-l)p_n(m-r) \nonumber\\
&=\delta_{k,n}\delta_{l,r}p_k(m-l)[1-p_k(m-l)]
\end{align}
Moreover, the reporting channel is characterized by 
\begin{equation}\label{E21} 
\omega_{\mathbf{z}}(k,n;l,r) = \sigma_z^2\delta_{k,n}\delta_{l,r}
\end{equation}
Note that $\boldsymbol{\theta}$, $\mathbf{u}$, and $\mathbf{z}$ are assumed to be independent of each other. We also need $\mathbf{p}(m) \triangleq \left [ p_1(m), ... , p_K(m) \right ]^T$  and $\mathbf{b}(m)\triangleq \left [ b_1(m), ... , b_K(m) \right ]^T$ in our formulations. $\mathbf{b}(m)$ denotes a realization of the random vector $\boldsymbol{\theta}(m) $. For simplicity, we will drop the time index $m$ when representing vectors and matrices. 

The reported sensing outcomes feed a linear estimator, depicted in Fig. \ref{fig1} as compensator, which uses spatial and temporal correlations of the received test summaries to estimate their actual (i.e., non-contaminated and non-interrupted) values. Specifically, the observation vector used by the estimator is $\mathbf{y}_{L}$ which, based on the notation provided in Section \ref{subsection Definitions}, can be expressed as
\begin{equation}\label{E22} 
\mathbf{y}_{L}=\boldsymbol{\tilde{\Theta}}_{L}\mathbf{u}_{L}+\mathbf{z}_{L}
\end{equation}
Note that $\boldsymbol{\tilde{\Theta}}_{L}$ is a $K(L+1)\times K(L+1)$ diagonal matrix. 
The estimation process at the FC is represented as
\begin{equation}\label{E23} 
 \mathbf{\hat{u}} = \boldsymbol{\xi}^T\mathbf{y}_{L}+\boldsymbol{\epsilon}
\end{equation}
where the weight vector $\boldsymbol{\xi}$ and the constant $\boldsymbol{\epsilon}$ are obtained as the minimizer of the mean-squared error (MMSE)
\begin{equation}\label{E24} 
 J\left ( \boldsymbol{\xi},\boldsymbol{\epsilon} \right )=E\left [ \left \|\mathbf{\hat{u}}-\mathbf{u}  \right \|^2 \right ]=E\left [ \left \|\boldsymbol{\xi}^T\mathbf{y}_{L}+\boldsymbol{\epsilon}-\mathbf{u}  \right \|^2 \right ].
\end{equation}
Accordingly, the MMSE weight vector can be obtained as
\begin{equation}\label{E25} 
 \boldsymbol{\xi}^*=\mathbf{C}_{\mathbf{y}_{L}}^{-1}\mathbf{C}_{\mathbf{y}_{L}\mathbf{u}}.
\end{equation}
Moreover, since $\mathbf{u}$ has nonzero mean in general, the optimum value of the bias term $\boldsymbol{\epsilon}$ is given by
\begin{equation}\label{E26} 
\boldsymbol{\epsilon}^* = E[\mathbf{u}] - \boldsymbol{\xi}^{*T} E[\mathbf{y}_{L}].
\end{equation}

The use of linear estimation---instead of e.g., maximum \emph{a posteriori} (MAP) estimation---in the proposed design is supported by several factors. First, to implement a linear estimator only first- and second-order statistics of the local sensing outcomes are required. Second, the analysis based on linear estimation leads to tractable formulations which facilitate performance optimization. And third, as it is shown in Section \ref{section Numerical Results}, the proposed structure achieves the optimal performance of  existing cooperative sensing schemes but with a significant level of energy efficiency.

From (\ref{E25}) and (\ref{E26}) we see that in order to establish the optimal linear estimation, we need $\mathbf{C}_{\mathbf{y}_{L}\mathbf{u}}$ and $\mathbf{C}_{\mathbf{y}_{L}}$. We derive these covariances in terms of the statistics of the local sensing outcomes and random interruptions as follows. First note that $\mathbf{C}_{\mathbf{y}_{L}\mathbf{u}}$ is  obtained as
 \begin{equation}\label{E27} 
\mathbf{C}_{\mathbf{y}_{L}\mathbf{u}}=\mathbf{\tilde{P}}_L \mathbf{C}_{\mathbf{u}_L \mathbf{u}}.
\end{equation}
Note that $\mathbf{\tilde{P}}_L = E[\boldsymbol{\tilde{\Theta}}_{L}]$. Then, by using (\ref{E19})--(\ref{E21}) the elements of $\mathbf{C}_{\mathbf{y}_{L}|\mathcal{H}_h}$ can be obtained as in (\ref{E28}) at the top of the following page. Recall from Section II-A that $k = \left \lceil \frac{i}{L+1}  \right \rceil$, $n = \left \lceil \frac{j}{L+1}  \right \rceil$, $l = (i+L)\textup{mod}(L+1)$, and $r = (j+L)\textup{mod}(L+1)$. Hence, we have shown that
\begin{figure*}[t!]
\normalsize
\begin{IEEEeqnarray}{rCl}\label{E28}
&\left (\mathbf{C}_{\mathbf{y}_L|\mathcal{H}_h} \right )_{i,j}  =  E\left[\left ( \theta_k(m-l)u_k(m-l)+ z_k(m-l)\right )\left (\theta_n(m-r)u_n(m-r)+ z_n(m-r)  \right )|\mathcal{H}_h\right]\nonumber\\
&-E\left[\left ( \theta_k(m-l)u_k(m-l)\right )|\mathcal{H}_h\right] E\left[\left ( \theta_n(m-r)u_n(m-r)\right )|\mathcal{H}_h\right] \nonumber\\
&=c_{\boldsymbol{\theta}}(k,n;l,r)\omega_{\mathbf{u}|\mathcal{H}_h}(k,n;l,r)+p_k(m-l)p_n(m-r) c_{\mathbf{u}|\mathcal{H}_h}(k,n;l,r)+\omega_{\mathbf{z}}(k,n;l,r)\nonumber\\
&=\delta_{k,n}\delta_{l,r}p_k(m-l)(1-p_k(m-l))\omega_{\mathbf{u}|\mathcal{H}_h}(k,k;l,l)+p_k(m-l)p_n(m-r) c_{\mathbf{u}|\mathcal{H}_h}(k,n;l,r)+\sigma_z^2\delta_{k,n}\delta_{l,r} \nonumber\\
&=\delta_{i,j}p_k(m-l)(1-p_k(m-l))\omega_{\mathbf{u}|\mathcal{H}_h}(k,k;l,l)+p_k(m-l)p_n(m-r) c_{\mathbf{u}|\mathcal{H}_h}(k,n;l,r)+\sigma_z^2\delta_{i,j}\nonumber\\
&=\left [\mathbf{\tilde{P}}_L\left ( \mathbf{I}-\mathbf{\tilde{P}}_L \right ) \boldsymbol{\mathfrak{R}}_{\mathbf{u_L}|\mathcal{H}_h}+\mathbf{\tilde{P}}_L \mathbf{C}_{\mathbf{u_L}|\mathcal{H}_h}\mathbf{\tilde{P}}_L + \sigma_z^2 \mathbf{I}  \right ]_{i,j}
\label{eq:floatingequation}
\end{IEEEeqnarray}
\hrulefill
\vspace*{0pt}
\end{figure*} 
\begin{equation}\label{E29} 
\mathbf{C}_{\mathbf{y}_L|\mathcal{H}_h} = \mathbf{\tilde{P}}_L\left ( \mathbf{I}-\mathbf{\tilde{P}}_L \right ) \boldsymbol{\mathfrak{R}}_{\mathbf{u_L}|\mathcal{H}_h}+\mathbf{\tilde{P}}_L \mathbf{C}_{\mathbf{u_L}|\mathcal{H}_h}\mathbf{\tilde{P}}_L + \sigma_z^2 \mathbf{I} 
\end{equation}
 and consequently,
\begin{equation}\label{E30} 
\mathbf{C}_{\mathbf{y}_L} = \mathbf{\tilde{P}}_L\left ( \mathbf{I}-\mathbf{\tilde{P}}_L \right ) \boldsymbol{\mathfrak{R}}_{\mathbf{u_L}}+\mathbf{\tilde{P}}_L \mathbf{C}_{\mathbf{u_L}}\mathbf{\tilde{P}}_L + \sigma_z^2 \mathbf{I}.
\end{equation}

It is worth noting that, $\mathbf{C}_{\mathbf{u}_L \mathbf{u}}$, $\mathbf{C}_{\mathbf{u_L}}$, and $\boldsymbol{\mathfrak{R}}_{\mathbf{u_L}}$ are obtained in terms of the second-order statistics of the channel gains between the PU and SUs as well as noise variances experienced by the sensors. The details are provided in Appendix I. Availability of these channel statistics is a common assumption in designing cooperative sensing schemes for CRNs \cite{Taricco11, Quan10, Quan08}. Since most PUs of interest such as current TV stations and OFDM systems operate based on certain standards, transmit at fixed power levels, and/or periodically transmit pilot signals, even exact channel gains can be obtained in many practical cases, see e.g., \cite{Paysarvi-Hoseini11}. 

According to the central limit theorem, if the number of samples $N$ in (\ref{E16}) is large enough, the local test summaries follow the Gaussian distribution. Consequently, the estimated local test summaries conditioned on a given realization of $\boldsymbol{\theta}_L$ are normal random variables, i.e., their conditional probability density function (pdf) can be represented as (for $h=0,1$)
\begin{align}\label{E31} 
&f\left ( {\hat{\mathbf{u}}}| \mathcal{H}_h, \boldsymbol{\theta}_L=\mathbf{b}_L \right )=\frac{1}{\sqrt{(2\pi)^K |\mathbf{C}_{\hat{\mathbf{u}}| \left \{ \mathcal{H}_h , \mathbf{b}_L  \right \}}   |}  } \nonumber\\
&\times \textup{exp}\left [ -\frac{1}{2}\left ( {\hat{\mathbf{u}}}-\boldsymbol{\mu}_{\hat{\mathbf{u}}| \left \{ \mathcal{H}_h , \mathbf{b}_L  \right \}}  \right ) \mathbf{C}_{\hat{\mathbf{u}}| \left \{ \mathcal{H}_h , \mathbf{b}_L  \right \}} ^{-1} \left ( {\hat{\mathbf{u}}}-\boldsymbol{\mu}_{\hat{\mathbf{u}}| \left \{ \mathcal{H}_h , \mathbf{b}_L  \right \}}  \right )^T\right ]
\end{align}
where 
\begin{align}\label{E32} 
\boldsymbol{\mu}_{\hat{\mathbf{u}}| \left \{ \mathcal{H}_h , \mathbf{b}_L  \right \}}&\triangleq E\left [ \hat{\mathbf{u}}|  \mathcal{H}_h , \boldsymbol{\theta}_L=\mathbf{b}_L  \right ] \\\nonumber
& =\boldsymbol{\xi}^{*T}E\left [ \boldsymbol{\tilde{\Theta}}_{L}\mathbf{u}_{L}+\mathbf{z}_{L}|  \mathcal{H}_h, \boldsymbol{\theta}_L=\mathbf{b}_L  \right ]+ \boldsymbol{\epsilon}^*\\\nonumber &=\boldsymbol{\xi}^{*T}\mathbf{\tilde{B}}_{L}E\left [ \mathbf{u}_{L}|  \mathcal{H}_h  \right ]+ \boldsymbol{\epsilon}^*
\end{align}
and $\mathbf{C}_{\hat{\mathbf{u}}| \left \{ \mathcal{H}_h , \mathbf{b}_L  \right \}}$ denotes the conditional covariance matrix of $\hat{\mathbf{u}}$. Note that, conditioned on $\boldsymbol{\theta}_L=\mathbf{b}_L $, the observation vector is $\mathbf{y}_{L}=\mathbf{\tilde{B}}_{L}\mathbf{u}_{L}+\mathbf{z}_{L}$ which leads to
\begin{equation}\label{E33} 
\mathbf{C}_{\mathbf{y}_L| \left \{ \mathcal{H}_h , \mathbf{b}_L  \right \}} = \mathbf{\tilde{B}}_{L}\mathbf{C}_{\mathbf{u}_L|\mathcal{H}_h}\mathbf{\tilde{B}}_{L}+\mathbf{C}_{\mathbf{z}_L}.
\end{equation}
 Therefore, by using (\ref{E23}) we have
\begin{align}\label{E34}
\mathbf{C}_{\hat{\mathbf{u}}| \left \{ \mathcal{H}_h , \mathbf{b}_L  \right \}}&=\boldsymbol{\xi}^{*T} \mathbf{C}_{\mathbf{y}_L| \left \{ \mathcal{H}_h , \mathbf{b}_L  \right \}}\boldsymbol{\xi}^*\\\nonumber
&=\boldsymbol{\xi}^{*T}\left (\mathbf{\tilde{B}}_{L}\mathbf{C}_{\mathbf{u}_L|\mathcal{H}_h}\mathbf{\tilde{B}}_{L}+\mathbf{C}_{\mathbf{z}_L}  \right )\boldsymbol{\xi}^*\\\nonumber
&=\boldsymbol{\xi}^{*T}\left (\mathbf{\tilde{B}}_{L}\mathbf{C}_{\mathbf{u}_L| \mathcal{H}_h}\mathbf{\tilde{B}}_{L}+\sigma_z^2\mathbf{I}  \right )\boldsymbol{\xi}^*
\end{align}
Now, the total probability theorem gives the distribution of the estimated local test summaries, i.e.,  
\begin{equation}\label{E35} 
f\left ({\hat{\mathbf{u}}}| \mathcal{H}_h \right )=\sum_{\mathbf{b}_L=0}^{2^n-1}f_{\hat{\mathbf{u}}}\left ( {\hat{\mathbf{u}}}| \mathcal{H}_h, \boldsymbol{\theta}_L=\mathbf{b}_L \right )\textup{Pr}\left \{ \boldsymbol{\theta}_L=\mathbf{b}_L \right \}
\end{equation}
where $n \triangleq K(L+1)$ and
\begin{align}\label{E36} 
\textup{Pr}&\left \{ \boldsymbol{\theta}_L=\mathbf{b}_L \right \} \nonumber \\
&= \prod_{l=0}^{L}\prod_{k=1}^{K}p_k(m-l)^{b_k(m-l)}[1-p_k(m-l)]^{1-b_k(m-l)}
\end{align}
Therefore, for a given $\mathbf{p}_L$ we can establish the optimal fusion rule at the FC as the well-known LRT, i.e., 
\begin{equation}\label{E37} 
\Lambda(\hat{\mathbf{u}})\triangleq\frac{f\left ( \hat{\mathbf{u}}| \mathcal{H}_1 \right )}{f\left ( \hat{\mathbf{u}}| \mathcal{H}_0 \right )} \begin{matrix}
\small{\mathcal{H}_1}\\ 
\gtrless\\ 
\small{\mathcal{H}_0} 
\end{matrix} \tau_{\Lambda}
\end{equation}
where the threshold $\tau_{\Lambda}$ is determined based on minimizing the average Bayesian cost of detection.

It is worth pointing out that, in existing cooperative sensing schemes  \cite{Taricco11, Quan10, Quan08} where there is no interruption nor is any compensation\footnote{Note that in this case $\hat{\mathbf{u}} = \mathbf{u} + \mathbf{z}$.}, the optimal performance of the LRT detection is closely achieved by a linear detector where the weighting vector is derived in terms of the listening-channel and reporting-channel statistics. More specifically, a so-called global test summary, which is used to decide the presence or absence of the PU signal, is constructed as a weighted sum of the received reported sensing outcomes. In the proposed system, the linear combining is represented as
\begin{equation}\label{E38} 
S(\hat{\mathbf{u}})\triangleq \mathbf{w}^T\hat{\mathbf{u}}
 \begin{matrix}
\small{\mathcal{H}_1}\\ 
\gtrless\\ 
\small{\mathcal{H}_0} 
\end{matrix} \tau_{S}
\end{equation}
where $S(\hat{\mathbf{u}})$ is the global test summary. 

The main idea of linear combining is that the combining weight for the signal from a particular user represents its contribution to the global decision. For example, if a CR generates a reliable (i.e., high-SNR) signal that may lead to correct detection on its own, it should be assigned a larger weighting coefficient. For those SUs experiencing deep fading or shadowing, their weights are decreased in order to reduce their negative contribution to the decision fusion. 

Although linear combining is known to be an effective and computationally-affordable technique in designing cooperative sensing schemes, it is not energy-efficient. It is worth noting that, regardless of the channel conditions experienced by different sensing nodes, they all provide their sensing outcomes at the same cost. This cost manifests itself in the form of energy consumed at the local sensing and reporting phases. When the contribution of a particular node to the global decision is suppressed by assigning a small weight to it, the energy consumed by that node for the local sensing and reporting processes is relatively wasted. 

In the proposed design, instead of suppressing the contribution of nodes working under deep fading or shadowing, they are occasionally ordered not to cooperate. In this structure, the more reliable a sensor is, the more likely it is to contribute to the overall sensing process. Consequently, the required discrimination between the reliable and unreliable sensors is achieved while no energy is wasted. Moreover, it is possible to optimize this process considering a constraint on the energy consumed at the local sensing and reporting phases. More specifically, we are interested in the optimal distribution associated with the random number generators, i.e., optimal value of $\mathbf{p}_L$, assuming a general linear fusion rule at the FC.    

\emph{Remark 1}: The random numbers are not required to be sent to the CR nodes during the sensing process. Hence, there is no intervention required from the FC while the local sensing is performed by the CR nodes. In fact, communication of the random numbers takes place before starting the sensing process. Based on the listening-channel and reporting-channel statistics, the distribution of the binary random numbers are first calculated at the FC and then, each CR node can either receive a random bit string generated according to the corresponding Bernoulli distribution, or receive only two numbers to feed a simple binary random number generator. One of the numbers indicates the corresponding Bernoulli distribution and the other one serves as a seed for the local random number generator. Once initialized, the cooperative sensing is performed in a distributed manner without any intervention from the FC. We know that in practical CRN architectures, such as the one assumed in the IEEE 802.22 standard \cite{Cordeiro06}, the FC regularly communicates with the CR nodes to coordinate their communication-related activities as well as their local sensing and reporting processes. Therefore, the proposed cooperation structure fits easily to the commonly-used CRN configurations.

\section{Problem Formulation}\label{section NPC}
In binary hypothesis testing for signal detection, the detector performance is commonly characterized by two probabilities, namely, the probability of false alarm $P_{\textup{f}}$ and the probability of detection $P_{\textup{d}}$. 
The system performance optimization based on NPC requires maximizing the probability of detection subject to a constant false alarm probability, i.e.,
\begin{equation}\label{E39} 
\operatorname*{max}P_{\textup{d}} ~~\textup{s.t.}~~ P_{\textup{f}}=\alpha
\end{equation}
where $\alpha$ is referred to as the significance level of the detector.

Detection and false alarm probabilities depend on the global test summary statistics. Since $\mathbf{u}_L$ and $\mathbf{z}_L$ follow normal distributions, we can conclude from (\ref{E22}), (\ref{E23}), and (\ref{E38}) that, conditioned on $\boldsymbol{\theta}_L = \mathbf{b}_L $, our test summary $S(\hat{\mathbf{u}})$ is a normal random variable with the following mean and variance (for $h=0,1$)
\begin{align}\label{E4142} 
E\left [ S(\hat{\mathbf{u}})|\mathcal{H}_h, \boldsymbol{\theta}_L = \mathbf{b}_L \right ]&=\mathbf{w}^T\boldsymbol{\mu}_{\hat{\mathbf{u}}| \left \{ \mathcal{H}_h , \mathbf{b}_L\right \} }\\
\textup{Var}\left [ S(\hat{\mathbf{u}})|\mathcal{H}_h, \boldsymbol{\theta}_L = \mathbf{b}_L \right ]&=\mathbf{w}^T\mathbf{C}_{\hat{\mathbf{u}}| \left \{ \mathcal{H}_h , \mathbf{b}_L  \right \}}\mathbf{w}
\end{align}
Consequently, the detection and false alarm probabilities, conditioned on $\boldsymbol{\theta}_L = \mathbf{b}_L$, can be expressed as
\begin{align}\label{E43}
P_{\textup{d}|\mathbf{b}_L} &\triangleq \textup{Pr} \left \{S>\tau_S|\mathcal{H}_1, \boldsymbol{\theta}_L = \mathbf{b}_L  \right \}\\\nonumber 
&=Q\left ( \frac{\tau_S-\mathbf{w}^T\boldsymbol{\mu}_{\hat{\mathbf{u}}| \left \{ \mathcal{H}_1 , \mathbf{b}_L  \right \}}}{\sqrt{\mathbf{w}^T\mathbf{C}_{\hat{\mathbf{u}}| \left \{ \mathcal{H}_1 , \mathbf{b}_L  \right \}}\mathbf{w}}} \right ) 
\end{align}
\begin{align}\label{E44}
P_{\textup{f}|\mathbf{b}_L} &\triangleq \textup{Pr} \left \{S>\tau_S|\mathcal{H}_0, \boldsymbol{\theta}_L = \mathbf{b}_L  \right \}\\\nonumber 
&=Q\left ( \frac{\tau_S-\mathbf{w}^T\boldsymbol{\mu}_{\hat{\mathbf{u}}| \left \{ \mathcal{H}_0 , \mathbf{b}_L  \right \}}}{\sqrt{\mathbf{w}^T\mathbf{C}_{\hat{\mathbf{u}}| \left \{ \mathcal{H}_0 , \mathbf{b}_L  \right \}}\mathbf{w}}} \right )  
\end{align}
where $P_{\textup{f}|\mathbf{b}_L}$ and $P_{\textup{d}|\mathbf{b}_L}$ denote the false alarm and detection probabilities conditioned on $\boldsymbol{\theta}_L = \mathbf{b}_L$, respectively. 
The total probability theorem then gives the overall detection and false alarm probabilities based on their conditional counterparts, i.e.,  
\begin{eqnarray}
P_{\textup{d}} = \sum_{\mathbf{b}_L = 0}^{2^n-1}  P_{\textup{d}|\mathbf{b}_L} \textup{Pr}\left \{  \boldsymbol{\theta}_L = \mathbf{b}_L \right \} \label{E45} \\
P_{\textup{f}} = \sum_{\mathbf{b}_L = 0}^{2^n-1}  P_{\textup{f}|\mathbf{b}_L} \textup{Pr}\left \{  \boldsymbol{\theta}_L = \mathbf{b}_L \right \} \label{E46} 
\end{eqnarray}

Given the target false alarm probability $P_{\textup{f}} = \alpha$, the detection probability can be obtained by removing $\tau_S$ from \eqref{E43} and \eqref{E44}. More specifically, solving $P_{\textup{f}|\mathbf{b}_L} = \alpha$ for $\tau_S$ gives
\begin{equation}\label{E47} 
\tau_S = Q^{-1}(\alpha)\sqrt{\mathbf{w}^T\mathbf{C}_{\hat{\mathbf{u}}| \left \{ \mathcal{H}_0 , \mathbf{b}_L  \right \}}\mathbf{w}}+\mathbf{w}^T\boldsymbol{\mu}_{\hat{\mathbf{u}}| \left \{ \mathcal{H}_0 , \mathbf{b}_L  \right \}}
\end{equation}
According to \eqref{E47}, having a constant false alarm probability requires a detection threshold which depends on the random number generator outcomes. In other words, unlike in \cite{Taricco11, Quan10, Quan08}, the detection threshold for a given false alarm probability in the proposed structure is not fixed and in fact, can be viewed as a random variable determined as a function of $ \boldsymbol{\theta}_L$. The proposed threshold adaptation is similar to the adaptation process in radar receivers with constant false alarm probability, a.k.a., constant false alarm rate (CFAR) systems. The difference here is that the threshold depends not only on the received noise level, but also on the set of nodes participating in the cooperative sensing.  

The detection threshold $\tau_S$, when plugged into \eqref{E43}, leads to the following conditional detection probability
\begin{equation}\label{E48} 
P_{\textup{d}|\mathbf{b}_L}^{(\alpha)} = Q\left ( \frac{Q^{-1}(\alpha)\sqrt{\mathbf{w}^T\mathbf{C}_{\hat{\mathbf{u}}| \left \{ \mathcal{H}_0 , \mathbf{b}_L  \right \}}\mathbf{w}}-\mathbf{a}_{\mathbf{b}_L}^T\mathbf{w}}{\sqrt{\mathbf{w}^T\mathbf{C}_{\hat{\mathbf{u}}| \left \{ \mathcal{H}_1 , \mathbf{b}_L  \right \}}\mathbf{w}}} \right )
\end{equation}
where $\textup{P}_{\textup{d}|\mathbf{b}_L}^{(\alpha)}$ denotes the conditional detection probability when the system false alarm probability is set to $\alpha$, and
\begin{align}\label{E49} 
\mathbf{a}_{\mathbf{b}_L}&\triangleq \boldsymbol{\mu}_{\hat{\mathbf{u}}| \left \{ \mathcal{H}_1 , \mathbf{b}_L  \right \}}-\boldsymbol{\mu}_{\hat{\mathbf{u}}| \left \{ \mathcal{H}_0 , \mathbf{b}_L  \right \}} \nonumber\\
&=\boldsymbol{\xi}^{*T}\mathbf{\tilde{B}}_{L}\left (E\left [ \mathbf{u}_{L}|  \mathcal{H}_1  \right ] -E\left [ \mathbf{u}_{L}|  \mathcal{H}_0  \right ] \right )\nonumber\\
&=\mathbf{C}_{\mathbf{u}_L \mathbf{u}}^T\mathbf{\tilde{P}}_L \left [ \mathbf{\tilde{P}}_L\left ( \mathbf{I}-\mathbf{\tilde{P}}_L \right ) \boldsymbol{\mathfrak{R}}_{\mathbf{u_L}}+\mathbf{\tilde{P}}_L \mathbf{C}_{\mathbf{u_L}}\mathbf{\tilde{P}}_L + \sigma_z^2 \mathbf{I} \right ]^{-1}\nonumber\\
&\times\mathbf{\tilde{B}}_{L}\left (E\left [ \mathbf{u}_{L}|  \mathcal{H}_1  \right ] -E\left [ \mathbf{u}_{L}|  \mathcal{H}_0  \right ] \right )
\end{align}
Moreover, by using \eqref{E25}, \eqref{E29}, and \eqref{E34} we have
\begin{align}\label{E50}
&\mathbf{C}_{\hat{\mathbf{u}}| \left \{ \mathcal{H}_h , \mathbf{b}_L  \right \}}=\nonumber\\
& \mathbf{C}_{\mathbf{u}_L \mathbf{u}}^T\mathbf{\tilde{P}}_L \left [ \mathbf{\tilde{P}}_L\left ( \mathbf{I}-\mathbf{\tilde{P}}_L \right ) \boldsymbol{\mathfrak{R}}_{\mathbf{u_L}}+\mathbf{\tilde{P}}_L \mathbf{C}_{\mathbf{u_L}}\mathbf{\tilde{P}}_L + \sigma_z^2 \mathbf{I} \right ]^{-1}\nonumber\\
&\times\left (\mathbf{\tilde{B}}_{L}\mathbf{C}_{\mathbf{u}_L| \mathcal{H}_h}\mathbf{\tilde{B}}_{L}+\sigma_z^2\mathbf{I}  \right )\nonumber \\
&\times\left [ \mathbf{\tilde{P}}_L\left ( \mathbf{I}-\mathbf{\tilde{P}}_L \right ) \boldsymbol{\mathfrak{R}}_{\mathbf{u_L}}+\mathbf{\tilde{P}}_L \mathbf{C}_{\mathbf{u_L}}\mathbf{\tilde{P}}_L + \sigma_z^2 \mathbf{I} \right ]^{-1}\mathbf{\tilde{P}}_L \mathbf{C}_{\mathbf{u}_L \mathbf{u}}
\end{align}

In order to clearly see the role of $\mathbf{p}_L$ in the proposed system performance, we use the fact that the elements of $\mathbf{\tilde{P}}_L$ lie between zero and one. Accordingly, we apply approximations on \eqref{E49} and \eqref{E50} based on first-order Taylor series expansion, i.e., $\left (\mathbf{I} + \mathbf{A}  \right )^{-1} \approx \mathbf{I} - \mathbf{A}$ when $\mathbf{A}^k \rightarrow \mathbf{0}_n $ as $k$ grows. As the result, we obtain 
\begin{align}\label{E5152} 
\mathbf{a}_{\mathbf{b}_L}&\approx \mathbf{C}_{\mathbf{u}_L\mathbf{u}}^T\mathbf{\tilde{P}}_{L}\mathbf{\tilde{B}}_{L}\left (E\left [ \mathbf{u}_{L}|  \mathcal{H}_1  \right ] -E\left [ \mathbf{u}_{L}|  \mathcal{H}_0  \right ] \right )\\
\mathbf{C}_{\hat{\mathbf{u}}| \left \{ \mathcal{H}_h , \mathbf{b}_L  \right \}}&\approx \mathbf{C}_{\mathbf{u}_L\mathbf{u}}^T\mathbf{\tilde{P}}_{L}\left (\mathbf{\tilde{B}}_{L}\mathbf{C}_{\mathbf{u}_L| \mathcal{H}_h}\mathbf{\tilde{B}}_{L}+\sigma_z^2\mathbf{I}  \right )\mathbf{\tilde{P}}_{L}\mathbf{C}_{\mathbf{u}_L\mathbf{u}}
\end{align}

Through some further algebraic manipulations, the conditional detection probability is reformulated as
\begin{equation}\label{E53} 
P_{\textup{d}|\mathbf{b}_L}^{(\alpha)}(\mathbf{p}_L) \approx Q\left ( \frac{Q^{-1}(\alpha)\sqrt{\mathbf{p}_L^T\boldsymbol{\Sigma}_{\left \{ \mathcal{H}_0 , \mathbf{b}_L  \right \}}\mathbf{p}_L}-\mathbf{q}_{\mathbf{b}_L}^T\mathbf{p}_L}{\sqrt{\mathbf{p}_L^T\boldsymbol{\Sigma}_{\left \{ \mathcal{H}_1 , \mathbf{b}_L  \right \}}\mathbf{p}_L}} \right )
\end{equation}
where, 
\begin{align}\label{E5455} 
\mathbf{q}_{\mathbf{b}_L}&\triangleq\mathbf{C}_{\mathbf{u}_L\mathbf{u}}\mathbf{w}\circ\mathbf{\tilde{B}}_{L}\left (E\left [ \mathbf{u}_{L}|  \mathcal{H}_1  \right ] -E\left [ \mathbf{u}_{L}|  \mathcal{H}_0  \right ] \right )\\
\boldsymbol{\Sigma}_{\left \{ \mathcal{H}_h , \mathbf{b}_L\right \}} &\triangleq \left (\mathbf{\tilde{B}}_{L}\mathbf{C}_{\mathbf{u}_L| \mathcal{H}_h}\mathbf{\tilde{B}}_{L}+\mathbf{I}  \right )\circ  \left (\mathbf{C}_{\mathbf{u}_L\mathbf{u}}\mathbf{w}\mathbf{w}^T\mathbf{C}_{\mathbf{u}_L\mathbf{u}}^T  \right )
\end{align}
For simplicity, we have assumed that $\sigma_z^2 = 1$. It can be easily shown that $\mathbf{\Sigma}_{\left \{ \mathcal{H}_h , \mathbf{b}_L\right \}}$ is positive semidefinite and the elements of $\mathbf{q}_{\mathbf{b}_L}$ are all nonnegative. 

From \eqref{E53} the overall detection probability for the target false alarm probability of $\alpha$ can be obtained as
\begin{eqnarray}\label{E56} 
P_{\textup{d}}^{(\alpha)} (\mathbf{p}_L)= \sum_{\mathbf{b}_L = 0}^{2^n-1}  \textup{P}_{\textup{d}|\mathbf{b}_L}^{(\alpha)} \textup{Pr}\left \{  \boldsymbol{\theta}_L = \mathbf{b}_L \right \}
\end{eqnarray}
where $P_{\textup{d}}^{(\alpha)}$ denotes the overall detection probability when the system exhibits a constant false alarm probability of $\alpha$. 

Based on the analysis provided, now we can represent our proposed detector performance optimization as
\begin{gather*}\label{P1}
\mathbf{p}_L^* = \operatorname*{argmax}_{\mathbf{p}_{L}} P_{\textup{d}}^{(\alpha)} (\mathbf{p}_L) \tag{P1} \\ 
\begin{matrix}
\textup{s.t.} &\mathbf{0}\preceq \mathbf{p}_L \preceq \mathbf{1}, & \mathbf{1}^T \cdot \mathbf{p}_L \leq (1-\eta)n
\end{matrix}
\end{gather*}
where $\eta$ denotes the power-efficiency constraint we consider in designing the proposed interrupted reporting scheme.  Note that for $0 < \eta \le 1$ some elements of $\mathbf{p}_L$ have to be less than 1, which means we are forcing some sensing nodes to occasionally go to the sleep mode (i.e., avoid cooperation). Moreover, it is clear that in general, the higher $\eta$ we chose, the higher is the number of nodes which are forced to sleep, or the more likely any node is to go to the sleep mode. In fact, $\eta$ is the parameter by which we control the energy consumption of the proposed interrupted reporting scheme. 

\emph{Remark 2}: Compared to the cooperation method in \cite{Taricco11, Quan10, Quan08}, the proposed approach consumes less energy. Specifically, in \cite{Taricco11, Quan10, Quan08} before starting the sensing process, a set of weights are calculated to discriminate between the sensing nodes whereas in the proposed method a set of Bernoulli probabilities are calculated with the same objective. Since we have not changed any of the major phases in the general cooperation process considered in those works, it can be concluded that, considering the entire sensing process, the proposed cooperation scheme is more efficient. 

Our optimization in \eqref{P1} is a challenging problem since, firstly, as shown by \eqref{E56}, the objective function is constructed as a summation over $2^n$ nonlinear functions of $\mathbf{p}_L$, and secondly, the summand functions are not concave. Solving such an optimization which needs exponentially-growing number of operations to be built would be a serious challenge even if all the summand functions were concave. In the following section, we tackle these issues by developing a stochastic optimization approach for the problem through considering \eqref{P1} as a decision making process with uncertainty. This approach will lead us to construct an effective low-complexity solution for the performance optimization of the proposed system in Section \ref{section DC}.  

\section{Stochastic Optimization}\label{section SO}
In this section, we develop a multistage stochastic programming approach based on a so-called \emph{scenario tree} structure \cite{Defourny11}. We first use this scenario tree to represent an alternative formulation of the optimization problem. This new formulation enables us to derive an alternative problem formulation whose objective function is built more conveniently. Next, we exploit the scenario tree to construct a two-stage stochastic optimization scheme which can serve as a fast procedure in the proposed system to obtain an approximate solution. The decision making structure developed in this section along with the two-stage stochastic optimization are used in Section \ref{section DC} to develop, based on standard semidefinite programming techniques, a polynomial-complexity optimization approach which gives very good results. 

\begin{figure}[]
	\centering 
  \includegraphics[scale=0.49]{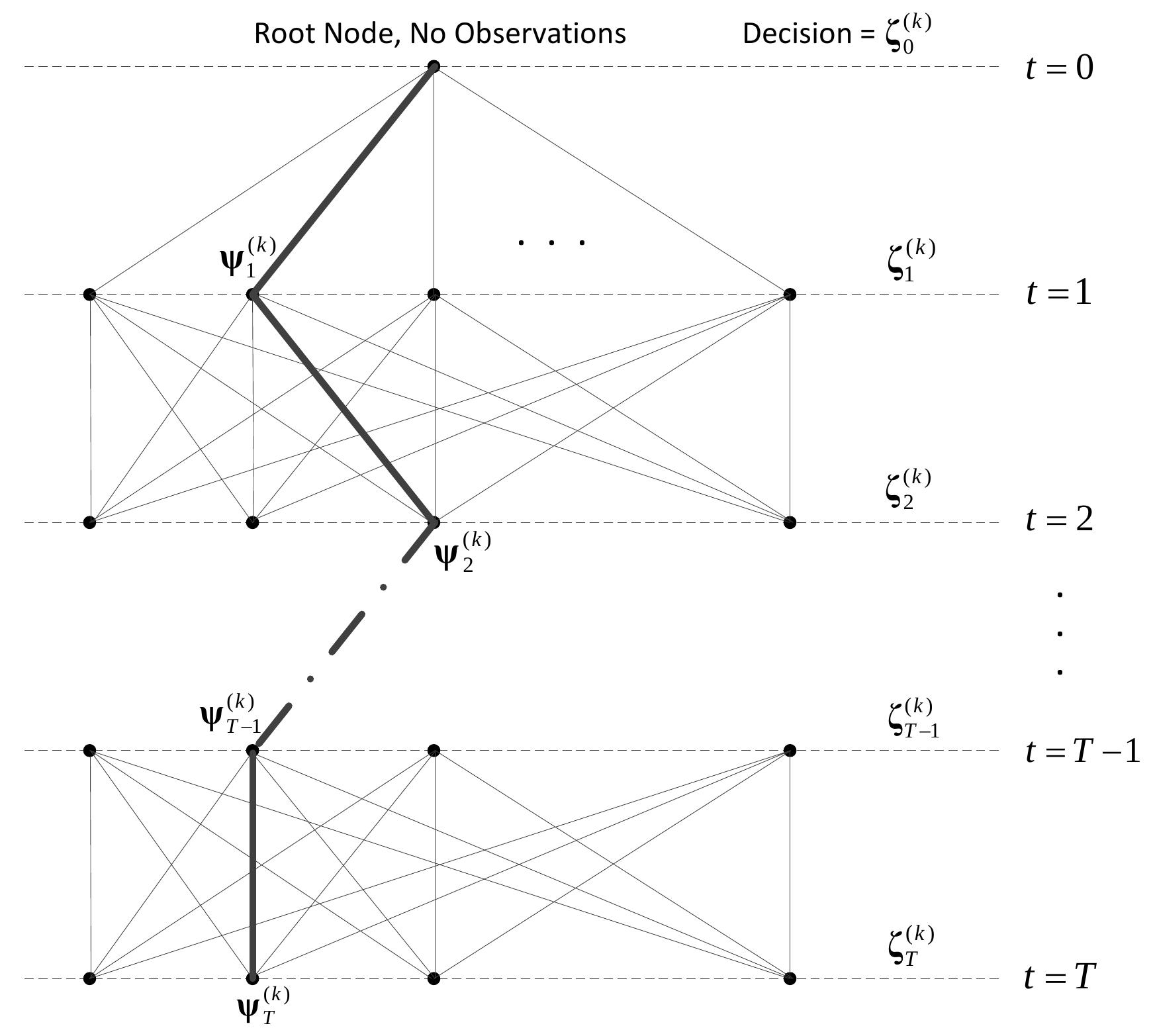} 
  \caption{Scenario tree for the proposed T-stage stochastic optimization.}
  \label{fig3-2}
\end{figure}

First note that the objective function in \eqref{P1} can be viewed as the expectation of the performance metric (i.e., the probability of detection) over the probability distribution of $\boldsymbol{\theta}_L$, i.e., 
\begin{eqnarray}\label{eqR330} 
\textup{P}^{(\alpha)}_{\textup{d}}(\mathbf{p}_L)=E_{\boldsymbol{\theta}_L}\left [\textup{P}^{(\alpha)}_{\textup{d}|\boldsymbol{\theta}_L}  \right ] 
\end{eqnarray}
From the stochastic programming point of view, $\boldsymbol{\theta}_L$ is the source of uncertainty in our decision making process. Hence, successive realizations of $\boldsymbol{\theta}_L$ form the branches of the scenario-tree considered, see Fig \ref{fig3-2}. 


The proposed scenario tree is constructed as follows. A \emph{root node} is associated to the first \emph{decision stage} and to the initial absence of observations. The first decision stage is designated as $t=0$. The nodes associated with decision stage $t$, $t=1,...,T$, are connected as children nodes to the nodes in stage $t-1$. The observation at the decision stage $t$ is denoted by $\boldsymbol{\psi}_t$ and defined as 
\begin{eqnarray}\label{SO57} 
\boldsymbol{\psi}_t \triangleq \mathbf{b}_L(m+(t-1)(L+1)) 
\end{eqnarray}
Each node at stage $t$ corresponds to one possible outcome for $\boldsymbol{\psi}_t$ given the observation  corresponding to its parent, i.e., $\boldsymbol{\psi}_{t-1}$. The observations associated to the nodes on the unique path from the root to a leaf define together a particular \emph{scenario}. Specifically, the $k$th scenario is denoted by $\boldsymbol{\psi}^{(k)} \triangleq [\boldsymbol{\psi}_1^{(k)}, ..., \boldsymbol{\psi}_T^{(k)}]$, which is associated with the $k$th branch of the scenario tree while $\boldsymbol{\psi}_t^{(k)}$ represents the observation on the $k$th branch at stage $t$.

At stage $t$ of the $k$th branch and after observing $\boldsymbol{\psi}_t^{(k)}$, a decision is made, denoted by $\boldsymbol{\zeta}^{(k)}_t$, which determines the Bernoulli distribution of the random variables for the next stage, i.e., for $t = 0, 1, ..., T$, we set
\begin{eqnarray}\label{SO58} 
 \mathbf{p}_L(m+t(L+1)) \coloneqq \boldsymbol{\zeta}^{(k)}_t
\end{eqnarray}
Recall that $\mathbf{p}_L(m) = \textup{Pr}\{\boldsymbol{\theta}_L(m) = \mathbf{b}_L(m)\}$.  Moreover, we know that each realization of $\boldsymbol{\theta}_L$ determines the state of the sensing nodes for $L+1$ consecutive time slots. Hence, the indexing distance between two consecutive decision stages should be $(L+1)$ as indicated in \eqref{SO57} and \eqref{SO58}. Note also that, the initial decision $\boldsymbol{\zeta}^{(k)}_0$ is made with no observations available.

We collect all the decisions associated with the $k$th scenario in $\boldsymbol{\zeta}^{(k)} \triangleq [\boldsymbol{\zeta}_0^{(k)}, \boldsymbol{\zeta}_1^{(k)}, ..., \boldsymbol{\zeta}_T^{(k)}]$. Based on these decisions, the occurrence probability of the $k$th scenario, denoted by $\pi^{(k)}$, is derived as
\begin{align}\label{eqR332} 
&\pi^{(k)} \triangleq \textup{Pr} \left \{ \boldsymbol{\psi}^{(k)} \right \} \nonumber \\
&= \textup{Pr} \left \{ \boldsymbol{\psi}_1^{(k)} \right \}\textup{Pr} \left \{ \boldsymbol{\psi}_2^{(k)}|\boldsymbol{\psi}_1^{(k)} \right \}\cdots \textup{Pr} \left \{ \boldsymbol{\psi}_T^{(k)}|\boldsymbol{\psi}_{T-1}^{(k)}, \cdots ,\boldsymbol{\psi}_1^{(k)} \right \} \nonumber \\
&= \varphi\left (\boldsymbol{\psi}_1^{(k)},\boldsymbol{\zeta}_0^{(k)}  \right )\varphi\left (\boldsymbol{\psi}_2^{(k)},\boldsymbol{\zeta}_1^{(k)}  \right ) \cdots \varphi\left (\boldsymbol{\psi}_T^{(k)},\boldsymbol{\zeta}_{T-1}^{(k)}  \right )
\end{align}
where $\varphi\left (\boldsymbol{\psi}_t^{(k)},\boldsymbol{\zeta}_{t-1}^{(k)}  \right )$ denotes the probability of observing $\boldsymbol{\psi}^{(k)} _t$ at stage $t$. This probability is calculated based on decided Bernoulli distributions as follows 
\begin{equation}\label{eqR333} 
\varphi\left (\boldsymbol{\psi}_t^{(k)},\boldsymbol{\zeta}_{t-1}^{(k)}  \right ) \triangleq \prod_{i=1}^{n}\boldsymbol{\zeta}_{t-1}^{(k)}(i)^{\boldsymbol{\psi}_{t}^{(k)}(i)}\left [1-\boldsymbol{\zeta}_{t-1}^{(k)}(i)  \right ]^{1-\boldsymbol{\psi}_{t}^{(k)}(i)}
\end{equation}
where index $i$ refers to the $i$th element of the vectors. 
 
Considering the $k$th scenario, based on the value of the random vector realized at each stage, i.e.,  $\boldsymbol{\psi}_t^{(k)}$, and Bernoulli distributions assigned, i.e., $\boldsymbol{\zeta}_{t}^{(k)}$, the probability of detection can be calculated, by using \eqref{E48}, as $\textup{P}_{\textup{d}|\boldsymbol{\psi}_t^{(k)}}^{(\alpha)}\left (\boldsymbol{\zeta}_{t}^{(k)}  \right )$. Consequently, the average probability of detection experienced on the $k$th branch of the scenario tree in $T$ stages is derived as
\begin{equation}\label{eqR333} 
\bar{\textup{P}}_{\textup{d}|\boldsymbol{\psi}^{(k)}}\left ( \boldsymbol{\zeta}^{(k)} \right ) \triangleq \frac{1}{T} \sum_{t=1}^{T}\textup{P}_{\textup{d}|\boldsymbol{\psi}_t^{(k)}}^{(\alpha)}\left (\boldsymbol{\zeta}_{t}^{(k)}  \right )
\end{equation}

We aim at maximizing the average probability of detection calculated over all possible scenarios. Therefore, based on the scenario tree developed,  we now represent our performance optimization in the following form
\begin{gather*}\label{SOP2}
\operatorname*{max}_{\left \{ \boldsymbol{\zeta}^{(k)}  \right \}_{k=1}^{N_{\textup{st}}} }\sum_{k=1}^{N_{\textup{st}}}\pi^{(k)}\bar{\textup{P}}_{\textup{d}|\boldsymbol{\psi}^{(k)}}\left ( \boldsymbol{\zeta}^{(k)} \right ) \tag{P2} \\ 
\textup{s.t.} ~~\mathbf{0}\preceq \boldsymbol{\zeta}_t^{(k)} \preceq \mathbf{1}~~\forall  k,t;\\
\mathbf{1}^T \cdot \boldsymbol{\zeta}_t^{(k)} \leq (1-\eta)n ~~ \forall  k,t;\\ 
\boldsymbol{\zeta}_0^{(k)} = \boldsymbol{\zeta}_0^{(j)}~~ \forall  k,j;\\
\boldsymbol{\zeta}_t^{(k)} = \boldsymbol{\zeta}_t^{(j)}~~\textup{whenever}\\
 ~[\boldsymbol{\psi}_1^{(k)}, ..., \boldsymbol{\psi}_{t-1}^{(k)}] \equiv [\boldsymbol{\psi}_1^{(j)}, ..., \boldsymbol{\psi}_{t-1}^{(j)}]. 
\end{gather*}
where $N_{\textup{st}}$ is the number of branches in the scenario tree. The last two constraints guarantee that the decision at each stage is \emph{non-anticipative}, which means that, it does not depend on the observations made at subsequent stages. It is clear that, the decisions cannot depend on observations that are not yet available. 

As a significant advantage of \eqref{SOP2} over \eqref{P1}, now we can decompose \eqref{SOP2} into smaller (deterministic) optimization problems. Specifically, since the $k$th summand in the objective function of  \eqref{SOP2} only depends on $\boldsymbol{\zeta}^{(k)}$, we can convert  \eqref{SOP2} into $N_{\textup{st}}$ smaller optimizations, i.e., for $k = 1, ..., N_{\textup{st}}$ we have 
\begin{gather*}\label{SOP3}
\operatorname*{max}_{ \boldsymbol{\zeta}^{(k)} }  \pi^{(k)}\bar{\textup{P}}_{\textup{d}|\boldsymbol{\psi}^{(k)}}\left ( \boldsymbol{\zeta}^{(k)} \right ) \tag{P3} \\ 
\textup{s.t.} ~~\mathbf{0}\preceq \boldsymbol{\zeta}_t^{(k)} \preceq \mathbf{1}~~\forall  k,t;\\
\mathbf{1}^T \cdot \boldsymbol{\zeta}_t^{(k)} \leq (1-\eta)n ~~ \forall  k,t;\\ 
\boldsymbol{\zeta}_0^{(k)} = \boldsymbol{\zeta}_0^{(j)}~~ \forall  k,j;\\
\boldsymbol{\zeta}_t^{(k)} = \boldsymbol{\zeta}_t^{(j)}~~\textup{whenever}\\
 ~[\boldsymbol{\psi}_1^{(k)}, ..., \boldsymbol{\psi}_{t-1}^{(k)}] \equiv [\boldsymbol{\psi}_1^{(j)}, ..., \boldsymbol{\psi}_{t-1}^{(j)}]. 
\end{gather*}
Note that \eqref{SOP2} is solved when all the $N_{st}$ optimizations in  \eqref{SOP3} are solved. Clearly, since $N_{st}$ grows exponentially with $n$, optimizing the system performance over the entire tree exhibits prohibitive complexity. However, decomposing the original optimization into a number of separate nonlinear programs enables us to propose a suboptimal approach for  \eqref{SOP2} with much lower computational burden in terms of the number of nonlinear programs to be solved. Specifically, instead of considering all the $N_{st}$ optimizations, one may chose to solve only the problem which corresponds to the branch that has just been observed. 
In this manner, the decisions are made based on an average of the performance metric estimated by using the observations in $T$ consecutive stages. 


Even dealing with one of the optimization problems in  \eqref{SOP3} is computationally demanding in the sense that, its objective function is obtained by summing over nonconcave functions. We develop an affordable suboptimal solution by simplifying the optimization as a two-stage stochastic program, which will serve as the foundation for a very effective optimization approach developed in Section \ref{section DC}. Specifically, we truncate the scenario tree with $T=1$ which leads to the following nonlinear optimization
\begin{gather*}\label{SOP4}
\operatorname*{max}_{\boldsymbol{\zeta}_1^{(k)}}\textup{P}_{\textup{d}|\boldsymbol{\psi}_1^{(k)}}^{(\alpha)}\left (\boldsymbol{\zeta}_{1}^{(k)}  \right )\tag{P4} \\ 
\textup{s.t.} ~~\mathbf{0}\preceq \boldsymbol{\zeta}_1^{(k)} \preceq \mathbf{1}\\
\mathbf{1}^T \cdot \boldsymbol{\zeta}_1^{(k)} \leq (1-\eta)n \\ 
\end{gather*}
which is equivalent to
%
%
%
%
\begin{gather*}\label{SOP5}
\mathbf{p}_L^* = \operatorname*{argmin}_{\mathbf{p}_{L}} \frac{Q^{-1}(\alpha)\sqrt{\mathbf{p}_L^T\boldsymbol{\Sigma}_{\left \{ \mathcal{H}_0 , \mathbf{b}_L  \right \}}\mathbf{p}_L}-\mathbf{q}_{\mathbf{b}_L}^T\mathbf{p}_L}{\sqrt{\mathbf{p}_L^T\boldsymbol{\Sigma}_{\left \{ \mathcal{H}_1 , \mathbf{b}_L  \right \}}\mathbf{p}_L}}  \tag{P5} \\ 
\begin{matrix}
\textup{s.t.} &\mathbf{0}\preceq \mathbf{p}_L \preceq \mathbf{1}, & \mathbf{1}^T \cdot \mathbf{p}_L \leq (1-\eta)n
\end{matrix}
\end{gather*}
Since the Q-function is monotonically decreasing, we have removed it form the objective function and turned the problem into a minimization. 

In the proposed cooperation scheme, for each different realization of the random interruptions we have a different set of cooperating nodes and consequently, we have a different detector. \eqref{SOP5} represents a decision making process in which for each $\mathbf{b}_L$ the value of $\mathbf{p}_L$  is specified by applying the NPC on the detector corresponding to $\mathbf{b}_L$. Therefore, in case \eqref{SOP5} is solved for each $\mathbf{b}_L$ generated by the random number generator, the Bernoulli distributions obtained are optimal for each realized cooperative detector in the NP sense. 
However, the decisions made by solving \eqref{SOP5} are obtained based on only one set of observations. In Section \ref{section DC}, we discuss  how to improve this decision making process by including more information about the effect of the random interruptions.

\eqref{SOP5} can be solved by applying the Karush-Kuhn-Tucker (KKT) conditions as follows. Note that the objective function in \eqref{SOP5} is homogeneous of degree zero with respect to $\mathbf{p}_L$. We first relax the constraints and minimize the objective function to obtain a lower bound on the optimal value of \eqref{SOP5}. Then, by using the homogeneity property of the objective function, we scale the derived point to find a feasible solution which attains this lower bound. More specifically, the lower bound is found by considering the Lagrange dual problem and applying the KKT conditions on the relaxed problem which yield \footnote{The lower bound can also be obtained by the quadratic programming approach proposed in \cite{Quan10}.}
\begin{equation}\label{E57}
\boldsymbol{\zeta}=\boldsymbol{\Sigma}_{\mathcal{H}_0}^{-1/2}\left [Q^{-1}(\alpha)\mathbf{I}+\kappa \mathbf{A}  \right ]^{-1}\mathbf{c}
\end{equation}
where $\boldsymbol{\zeta}$ is the global minimum of the relaxed problem, $\mathbf{A}\triangleq\boldsymbol{\Sigma}_{\mathcal{H}_1}\boldsymbol{\Sigma}_{\mathcal{H}_0}^{-1}$, and $\mathbf{c}\triangleq\boldsymbol{\Sigma}_{\mathcal{H}_0}^{-1/2}\mathbf{q}_{\mathbf{b}_L}$, while $\kappa$ is the root of the polynomial equation \cite{Taricco11}
\begin{equation}\label{E58}
\left \| \left [Q^{-1}(\alpha)\mathbf{I}+\kappa \mathbf{A}  \right ]^{-1}\mathbf{c} \right \|=1
\end{equation}
and satisfies 
\begin{equation}\label{E59}
Q^{-1}(\alpha)\mathbf{I}+\kappa \mathbf{A} > \mathbf{0}
\end{equation}
Then, the optimal solution of \eqref{SOP5} is derived as 
\begin{equation}\label{E60}
\mathbf{p}_L^*= \lambda \boldsymbol{\zeta}
\end{equation}
where
\begin{equation}\label{E61}
\lambda = \left\{\begin{matrix}
\frac{(1-\eta)n}{\mathbf{1}^T \boldsymbol{\zeta}}, & \textup{if} ~\frac{(1-\eta)n}{\mathbf{1}^T \boldsymbol{\zeta}} \boldsymbol{\zeta}\preceq \mathbf{1}\\ 
 \frac{\boldsymbol{\zeta}}{\operatorname*{max}_{i}\boldsymbol{\zeta}(i)},& \textup{otherwise}
\end{matrix}\right.
\end{equation}

\begin{figure}[]
	\centering 
  \includegraphics[scale=0.6]{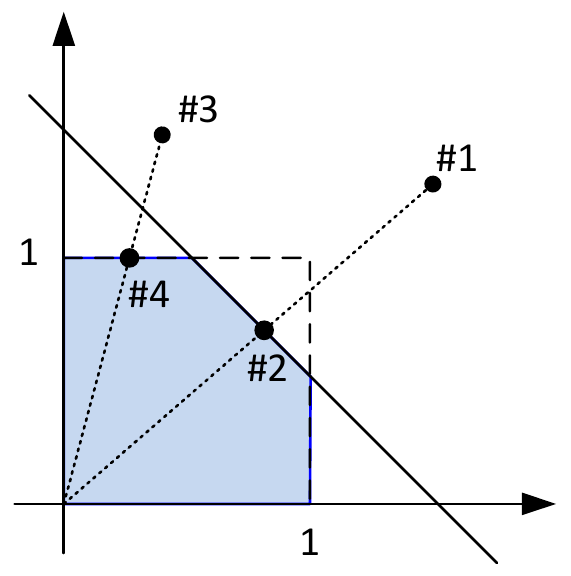} 
  \caption{Scaling $\boldsymbol{\zeta}$ to a feasible point of \eqref{SOP5}. Points \#1 and \#3 are mapped onto \#2 and \#4 respectively. The feasible point in each case is chosen to lie on the border of the feasibility region which is shown as a shaded area. }
  \label{fig3}
\end{figure}

Fig. \ref{fig3} demonstrates the scaling process indicated by \eqref{E60} and \eqref{E61}. In this figure, optimal points $\#2$ and $\#4$ are derived by multiplying points $\#1$ and $\#3$ with their respective values of $\lambda$. The feasible region of \eqref{SOP5} is depicted as the shaded area. As seen in this figure, $\lambda$ in \eqref{E61} maps $\boldsymbol{\zeta}$ to a point on the border of the feasibility region of \eqref{SOP5}. This particular scaling is justified by recalling that, we solve \eqref{SOP5} as a fast suboptimal approach for optimization of a system whose best performance is established originally as \eqref{P1}. Even though the optimal value of \eqref{SOP5} does not change by altering the norm of $\boldsymbol{\zeta}$, choosing a feasible point with the greatest norm leads to a better solution since the overall system performance depends on the norm of the decision vector $\mathbf{p}_L^*$  in general. 
\section{Deflection Criterion}\label{section DC}
Performance optimization based on deflection criterion means maximizing the DC which is defined as the variance-normalized distance between the centers of two conditional pdfs of the global test summary, i.e., 
\begin{equation}\label{E40} 
\Delta_m \triangleq \frac{E\left [S(\hat{\mathbf{u}})|\mathcal{H}_1  \right ]-E\left [S(\hat{\mathbf{u}})|\mathcal{H}_0  \right ] }{\sqrt{\textup{Var}\left \{ S(\hat{\mathbf{u}})| \mathcal{H}_0\right \}}} 
\end{equation}
In this section, we show the link between the two-stage stochastic programming \eqref{SOP5} and the DC. 

In the stochastic programming method developed in Section \ref{section SO}, the effect of the uncertainty is captured into the decision making process in the form of the statistics of the received reports conditioned on $\mathbf{b}_L$, i.e., $ \boldsymbol{\mu}_{\hat{\mathbf{u}}| \left \{ \mathcal{H}_h , \mathbf{b}_L  \right \}} $ and $\mathbf{C}_{\hat{\mathbf{u}}| \left \{ \mathcal{H}_h , \mathbf{b}_L  \right \}}$ (for $h = 0,1$).
We can improve this optimization approach by providing the decision maker with more information about the effect of $\boldsymbol{\theta}_L$. Specifically, we propose to make the decisions based on the  statistics $\boldsymbol{\mu}_{\hat{\mathbf{u}}| \mathcal{H}_h}$ and $\mathbf{C}_{\hat{\mathbf{u}}|  \mathcal{H}_h}$, rather than their $\mathbf{b}_L$-conditioned counterparts. Since the unconditioned statistics are the average of the conditional ones over $\boldsymbol{\theta}_L$, they provide the decision maker with more information about the impact of the interruptions. In consequence, by modifying \eqref{SOP5} we obtain the following optimization
\begin{gather*}\label{SOP6}
\mathbf{p}_L^* = \operatorname*{argmin}_{\mathbf{p}_{L}}\frac{Q^{-1}(\alpha)\sqrt{\mathbf{w}^T\mathbf{C}_{\hat{\mathbf{u}}| \mathcal{H}_0  }\mathbf{w}}-\mathbf{a}^T\mathbf{w}}{\sqrt{\mathbf{w}^T\mathbf{C}_{\hat{\mathbf{u}}|  \mathcal{H}_1 }\mathbf{w}}} \tag{P6} \\ 
\begin{matrix}
\textup{s.t.} &\mathbf{0}\preceq \mathbf{p}_L \preceq \mathbf{1}, & \mathbf{1}^T \cdot \mathbf{p}_L \leq (1-\eta)n
\end{matrix}
\end{gather*}
where 
\begin{align}\label{E62} 
\mathbf{a}&\triangleq \boldsymbol{\mu}_{\hat{\mathbf{u}}| \mathcal{H}_1 }-\boldsymbol{\mu}_{\hat{\mathbf{u}}| \mathcal{H}_0 }\nonumber 
 =\boldsymbol{\xi}^{*T}\mathbf{\tilde{P}}_L \left (E\left [\mathbf{u}_L|\mathcal{H}_1  \right ]-E\left [\mathbf{u}_L|\mathcal{H}_0  \right ]  \right ) \nonumber\\
& = \mathbf{C}_{\mathbf{y}_{L}\mathbf{u}}^T \mathbf{C}_{\mathbf{y}_{L}}^{-1}\mathbf{\tilde{P}}_L \left (E\left [\mathbf{u}_L|\mathcal{H}_1  \right ]-E\left [\mathbf{u}_L|\mathcal{H}_0  \right ]  \right )\nonumber\\
&\approx \mathbf{C}_{\mathbf{u}_L\mathbf{u}}^T \mathbf{\tilde{P}}_L^2 \left (E\left [\mathbf{u}_L|\mathcal{H}_1  \right ]-E\left [\mathbf{u}_L|\mathcal{H}_0  \right ]  \right )
\end{align}
and for $h=0,1$,
\begin{align}\label{E63}
&\mathbf{C}_{\hat{\mathbf{u}}|  \mathcal{H}_h }= \mathbf{C}_{\mathbf{y}_{L}\mathbf{u}}^T \mathbf{C}_{\mathbf{y}_{L}}^{-1}\mathbf{C}_{\mathbf{y}_{L}|\mathcal{H}_h}\mathbf{C}_{\mathbf{y}_{L}}^{-1}\mathbf{C}_{\mathbf{y}_{L}\mathbf{u}} \nonumber \\
&=\mathbf{C}_{\mathbf{u}_L \mathbf{u}}^T\mathbf{\tilde{P}}_L \left [ \mathbf{\tilde{P}}_L\left ( \mathbf{I}-\mathbf{\tilde{P}}_L \right ) \boldsymbol{\mathfrak{R}}_{\mathbf{u_L}}+\mathbf{\tilde{P}}_L \mathbf{C}_{\mathbf{u_L}}\mathbf{\tilde{P}}_L + \sigma_z^2 \mathbf{I} \right ]^{-1}\nonumber\\
&\times\left [ \mathbf{\tilde{P}}_L\left ( \mathbf{I}-\mathbf{\tilde{P}}_L \right ) \boldsymbol{\mathfrak{R}}_{\mathbf{u_L}|\mathcal{H}_h}+\mathbf{\tilde{P}}_L \mathbf{C}_{\mathbf{u_L}|\mathcal{H}_h}\mathbf{\tilde{P}}_L + \sigma_z^2 \mathbf{I} \right ]\nonumber \\
&\times\left [ \mathbf{\tilde{P}}_L\left ( \mathbf{I}-\mathbf{\tilde{P}}_L \right ) \boldsymbol{\mathfrak{R}}_{\mathbf{u_L}}+\mathbf{\tilde{P}}_L \mathbf{C}_{\mathbf{u_L}}\mathbf{\tilde{P}}_L + \sigma_z^2 \mathbf{I} \right ]^{-1}\mathbf{\tilde{P}}_L \mathbf{C}_{\mathbf{u}_L \mathbf{u}} \nonumber \\
&\approx  \mathbf{C}_{\mathbf{u}_L \mathbf{u}}^T\mathbf{\tilde{P}}_L^2 \mathbf{C}_{\mathbf{u}_L \mathbf{u}}
\end{align}
By using these approximations we see that the proposed optimization \eqref{SOP6} is almost equivalent to  
\begin{gather*}\label{SOP7}
\mathbf{p}_L^* = \operatorname*{argmax}_{\mathbf{p}_{L}}\frac{\mathbf{a}^T\mathbf{w}}{\sqrt{\mathbf{w}^T\mathbf{C}_{\hat{\mathbf{u}}|  \mathcal{H}_1 }\mathbf{w}}} \tag{P7} \\ 
\begin{matrix}
\textup{s.t.} &\mathbf{0}\preceq \mathbf{p}_L \preceq \mathbf{1}, & \mathbf{1}^T  \mathbf{p}_L \leq (1-\eta)n
\end{matrix}
\end{gather*}
Since $\mathbf{a}^T\mathbf{w}=E\left [S(\hat{\mathbf{u}})|\mathcal{H}_1  \right ]-E\left [S(\hat{\mathbf{u}})|\mathcal{H}_0  \right ]$ and $\mathbf{w}^T\mathbf{C}_{\hat{\mathbf{u}}|  \mathcal{H}_h }\mathbf{w}=\textup{Var} [S(\hat{\mathbf{u}})| \mathcal{H}_h]$, we clearly recognize the objective function in \eqref{SOP7} as the DC.  

Hence, we have shown that modifying the decision making process in \eqref{SOP5} to make the decisions based on overall (rather than conditional) statistics of the received interrupted reports is almost equivalent to optimizing the proposed system performance based on the deflection criterion. 
In addition, by the proposed modification the optimization variables derived are unpaired from $\boldsymbol{\theta}_L$. This means that unlike \eqref{SOP5}, this problem needs to be solved only once. 

Now we proceed with developing a solution for \eqref{SOP7}. 
By using the approximations in \eqref{E62} and \eqref{E63}, changing the variables, and with some further algebra we rewrite \eqref{SOP7} as 
\begin{gather*}\label{SOP8}
\boldsymbol{\pi}^* \triangleq\operatorname*{argmax}_{\boldsymbol{\pi}}\frac{\boldsymbol{\pi}^T\mathbf{D}\boldsymbol{\pi} }{\left \| \boldsymbol{\pi} \right \|} \tag{P8} \\ 
\begin{matrix}
\textup{s.t.} &\mathbf{0}\preceq \boldsymbol{\pi} \preceq \mathbf{D}_{\sigma}^{1/2}  \mathbf{1}, & \mathbf{1}^T \mathbf{D}_{\sigma}^{-1/2} \boldsymbol{\pi} \leq  (1-\eta)n
\end{matrix}
\end{gather*}
where $\boldsymbol{\pi} \triangleq \mathbf{D}_{\sigma}^{1/2}\mathbf{p_L}$ and $\mathbf{D} \triangleq \mathbf{D}_{\mu}\mathbf{D}_{\sigma}^{-1}$
while $\mathbf{D}_{\mu}$ and $\mathbf{D}_{\sigma}$ are diagonal matrices whose elements are derived as
\begin{equation}\label{eqR25} 
\left (\mathbf{D}_{\mu}  \right )_{i,j} \triangleq \delta_{i,j}\left [ \mathbf{C}_{\mathbf{u}_L 
\mathbf{u}}\mathbf{w}\left (E\left [\mathbf{u}_L|\mathcal{H}_1  \right ]-E\left [\mathbf{u}_L|\mathcal{H}_0  \right ] \right )^T \right ]_{i,j}
\end{equation}
\begin{equation}\label{eqR26} 
\left (\mathbf{D}_{\sigma}  \right )_{i,j}\triangleq \delta_{i,j}\left ( \mathbf{C}_{\mathbf{u}_L \mathbf{u}}\mathbf{w} \mathbf{w} ^T \mathbf{C}_{\mathbf{u}_L \mathbf{u}}^T\right)_{i,j}
\end{equation}

It is worth noting that, \eqref{SOP8} is a convex-over-convex fractional program over a convex set. This type of optimization is referred to as nonconcave fractional program and lacks a closed-form solution in general and is globally solved by using a special type of the BnB algorithm introduced in \cite{Benson06}. Although the BnB is known to be very effective for small- and moderate-size problems, its high (worst-case) computational complexity, which grows exponentially with the number of optimization variables, might make it difficult to apply when a large-scale network is considered.  

In order to derive a computationally-affordable solution, we convert \eqref{SOP8} into two nested optimization procedures. The inner procedure includes maximizing a QCQP and the outer procedure is realized as a one-dimensional optimization. Note that the feasible region of \eqref{SOP8} defines a polytope. The idea is to solve the problem on the intersection of the polytope with a ball centered at the origin, and to sweep the ball radius to scan the entire polytope searching for the global maximum. 
%
%

 We formulate the QCQP by introducing a new constraint to make the feasible region of \eqref{SOP8} be on a ball of radius $r$, $0 \le r \le  \sqrt{ \textup{Tr}\left (\mathbf{D}_{\sigma}\right )}$, and solve the following optimization for a constant $r$
\begin{gather*}\label{SOP9}
\varphi^*(r) \triangleq \frac{1}{r}\operatorname*{max}_{\boldsymbol{\pi}}\boldsymbol{\pi}^T\mathbf{D}\boldsymbol{\pi} \tag{P9} \\ 
\begin{matrix}
\textup{s.t.} &\mathbf{0}\preceq \boldsymbol{\pi} \preceq \mathbf{D}_{\sigma}^{1/2}  \mathbf{1}, & \mathbf{1}^T  \mathbf{D}_{\sigma}^{-1/2} \boldsymbol{\pi} \leq  (1-\eta)n,
\end{matrix}\\
\left\|\boldsymbol{\pi} \right\| = r
\end{gather*}
Then, the outer procedure, i.e., scanning the polytope, is realized as the following optimization 
\begin{gather*}\label{SOP10}
r^*\triangleq \operatorname*{argmax}_{r}\varphi^*(r)  \tag{P10} \\ 
\textup{s.t.}~ 0 \le r \le  \sqrt{ \textup{Tr}\left (\mathbf{D}_{\sigma}\right )}
\end{gather*}
Since \eqref{SOP10} is a one-dimensional problem, the computational complexity of the proposed nested loop is determined by \eqref{SOP9}. 
Solving \eqref{SOP9} requires maximizing a convex quadratic objective function over a convex set, which is NP-hard in general. Fortunately, numerous relaxation techniques are available in the literature for solving nonconvex quadratic programs. A variety of these relaxations can be formulated as semidefinite programs (SDPs) and it is shown in  \cite{Bao11} that the Shor relaxation \cite{shor90} when strengthened with the so-called reformulation-linearization technique (RLT) \cite{anstreicher09} dominates other commonly-used semidefinite relaxations with optimality gap averaging 3\%. 
Accordingly, we first reformulate \eqref{SOP9} by introducing a new variable as $\mathbf{V} = \boldsymbol{\pi}\boldsymbol{\pi}^T$ which leads to
\begin{gather*}\label{SOP9-a}
\operatorname*{max}_{\mathbf{V},\boldsymbol{\pi}}\textup{Tr}(\mathbf{D}\mathbf{V})  \tag{P9-a} \\ 
\textup{s.t.} 
\begin{cases}
\mathbf{0}\preceq \boldsymbol{\pi} \preceq \mathbf{D}_{\sigma}^{1/2} \cdot \mathbf{1}\\
\mathbf{1}^T  \mathbf{D}_{\sigma}^{-1/2} \boldsymbol{\pi} \leq  (1-\eta)n\\
\textup{Tr}(\mathbf{V}) = r^2\\
\mathbf{V} = \boldsymbol{\pi}\boldsymbol{\pi}^T
\end{cases}
\end{gather*}

The SDP relaxation is then realized by replacing the last constraint in \eqref{SOP9-a} by $\mathbf{V} \succeq \boldsymbol{\pi}\boldsymbol{\pi}^T$, while the RLT is constructed based on using products of upper and lower bound constraints on the original variables to obtain valid linear inequality constraints on the new variable $\mathbf{V}$. Consequently, the resulting SDP is
\begin{gather*}\label{SOP9-b}
\operatorname*{max}_{\mathbf{V},\boldsymbol{\pi}}\textup{Tr}(\mathbf{D}\mathbf{V})  \tag{P9-b} \\ 
\textup{s.t.} 
\begin{cases}
\mathbf{1}^T  \mathbf{D}_{\sigma}^{-1/2} \boldsymbol{\pi} \leq (1-\eta)n \\
\mathbf{V}-\boldsymbol{\pi}\mathbf{d}^{T}  - \mathbf{d} \boldsymbol{\pi}^{T} +\mathbf{d}\mathbf{d}^T  \ge 0 \\ 
\mathbf{V}-\boldsymbol{\pi}\mathbf{d}^{T}\le 0\\
\mathbf{V} \ge 0\\
\mathbf{V} \succeq \boldsymbol{\pi}\boldsymbol{\pi}^T
\end{cases}
\end{gather*}
where $\mathbf{d}  \triangleq \mathbf{D}_{\sigma}^{1/2} \mathbf{1} $. Note that the last constraint in this problem is equivalent to 
\begin{equation}\label{E66} 
\begin{pmatrix}
1 & \boldsymbol{\pi}^T \\ 
\boldsymbol{\pi} & \mathbf{V}
\end{pmatrix}\succeq 0
\end{equation}
Therefore, \eqref{SOP9-b} is a convex optimization problem which can be solved, to any arbitrary accuracy, in a numerically-reliable and efficient fashion.

This relaxation technique is known to provide very good approximate solutions for nonconvex quadratic programs in the form of \eqref{SOP9}. Nevertheless, \eqref{SOP9} can be solved for a global optimum in computational time complexity by a combinatorial optimization algorithm developed in \cite{Bienstock14}. Therefore, the proposed approach in converting the fractional program \eqref{SOP8} into a set of QCQPs is a practically-appealing approach even if the global optimum is concerned. In the following section, the effectiveness of the proposed method in optimizing the proposed system performance is verified by simulation results. 
 
\section{Results}\label{section Numerical Results}
In this section we visualize the performance of the proposed system by simulation results and make comparisons with the existing cooperative sensing schemes. In particular, we use the optimal linear combining scheme in \cite{Taricco11, Quan10, Quan08} as the benchmark.

First, we consider three sensing nodes, i.e., $K=3$, operating at different SNR regimes and  evaluate the detection and false alarm probabilities of the proposed system and compare them with the results predicted by our analysis. 

\begin{figure}[]
	\centering 
  \includegraphics[scale=0.30]{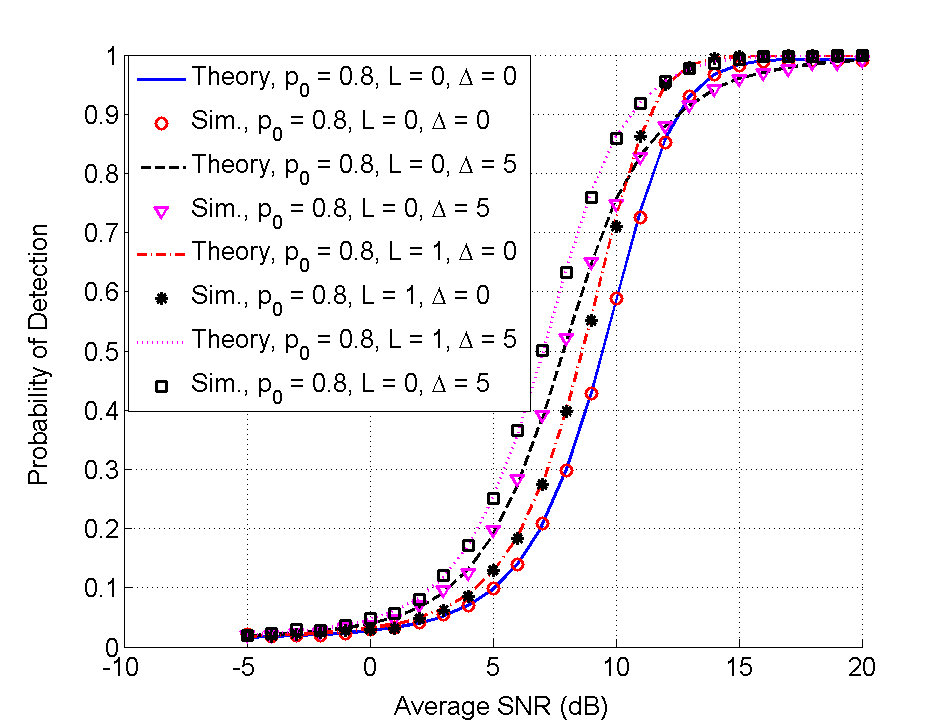} 
  \caption{The proposed system performance in different SNR regimes and with different SNR dispersion levels among the sensing nodes.}
  \label{fig4}
\end{figure}

Fig. \ref{fig4} depicts the detection probability under different average SNR levels and with different SNR dispersions among the sensing nodes. More specifically, the SNR levels considered are ${\textup{SNR}_0-\Delta, \textup{SNR}_0, \textup{SNR}_0+\Delta}$ where $\textup{SNR}_0$ represents the average SNRs among the sensing nodes and $\Delta$ denotes the SNR dispersion. The reporting channel noise variance is $\sigma_z^2 = 10$. The correlation coefficient $\rho$ between the local sensing outcomes  when $\mathcal{H}_1$ is true is $ 0.1$ and the number of samples $N$ used by each sensing node for energy detection is $20$. 
Spatio-temporal correlations between the local sensing outcomes are realized by applying a moving average filter on the local sensing results which are realized under the AWGN with fixed channel gains corresponding to the different SNR levels. The Bernoulli distributions are considered as $\mathbf{p}_L = p_0\mathbf{1}$ where $p_0$ indicates the average energy consumed for the local sensing and reporting purposes. For each set of theoretical and simulation results two values for SNR dispersion and two values for $L$ are considered. Hence we have eight curves in Fig. \ref{fig4}. The results in this figure are obtained for significance level $\alpha = 0.01$ and by 100,000 sample realizations for each average SNR level.  

It is clear from Fig. \ref{fig4}  that, regarding the probability of detection, the system performance predicted by the proposed analysis is in close agreement with the simulations. Moreover, we observe that the system performance is improved by taking into account the temporal correlations besides the spatial correlations. Specifically, the detection probability obtained in each case for $L = 1$ is higher than the performance obtained when $L = 0$. This is a reasonable observation as by using a more powerful compensation process, the degradations caused jointly by the interruption process and reporting channel contaminations are mitigated more effectively. 

\begin{figure}[]
	\centering 
  \includegraphics[scale=0.30]{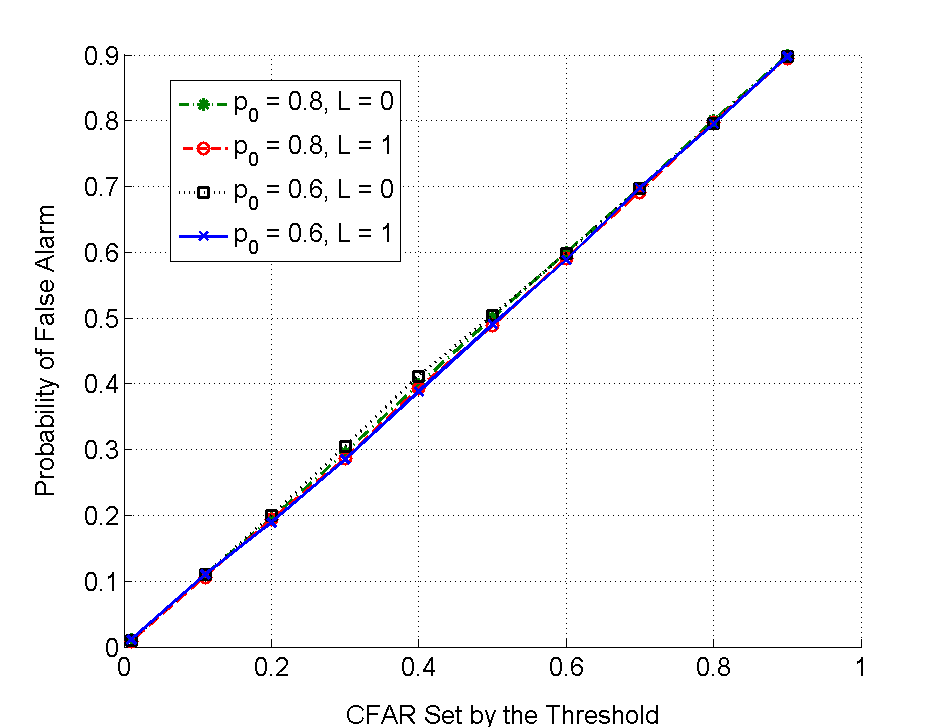} 
  \caption{Effect of the CFAR threshold set by \eqref{E47} on the false alarm rate of the proposed system.}
  \label{fig5}
\end{figure}

Analytical and simulation results concerning the false alarm probability of the proposed cooperative sensing are depicted in Fig. \ref{fig5}. This figure depicts the effect of the threshold adaptation scheme in \eqref{E47} on the false alarm probability (or more precisely, false alarm rate) observed. Specifically, in this figure, the false alarm rate corresponding to the adopted threshold level is compared against the false alarm rate that the system shows when using that threshold for detecting the PU signal. In this simulation, four cases are considered corresponding to $p_0 = 0.6, 0.8$ and $L = 0, 1$. We see in all these cases that, the proposed threshold adaptation works well, i.e., the observed false alarm rate of the detector is very close to the rate predicted by the analysis.  

\begin{figure}[]
	\centering 
  \includegraphics[scale=0.30]{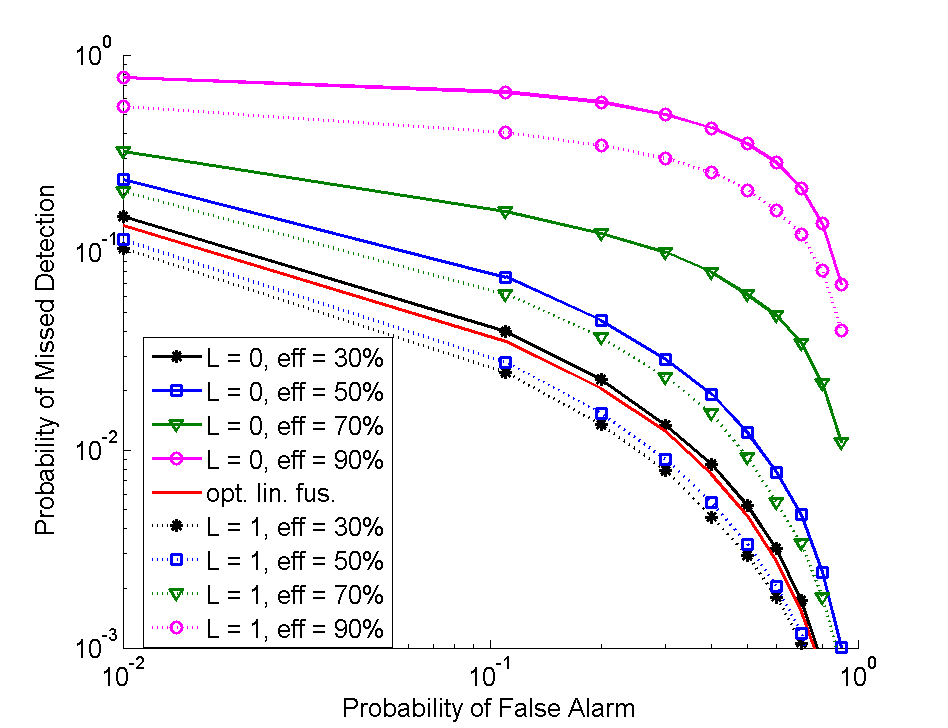} 
  \caption{CROC curves demonstrating the proposed system performance operating with different energy efficiency levels compared to the optimal linear fusion scheme.}
  \label{fig6}
\end{figure}

In Fig. \ref{fig6}, the system performance under different levels of energy efficiency is compared with the performance of the optimal linear combining scheme developed in \cite{Taricco11, Quan10, Quan08}. In this simulation, the complementary receiver operational characteristics (CROC) curves are plotted indicating the system performance in terms of the probability of false alarm and the probability of missed detection. The interrupted cooperation is simulated in two cases, i.e., for $L = 0$ and for $L = 1$, and in each case, four levels of energy efficiency are considered, i.e., efficiency levels of 30\%, 50\%, 70\%, and 90\% for three sensing nodes operating at SNR levels $\{12, 5, 10\}$ in dB. In this figure, we first observe the tradeoff between the energy efficiency and the detection quality. Specifically, the missed detection probability is increased when the system works with higher levels of energy efficiency. Moreover, for a given efficiency level, the proposed system exhibits a better performance for $L = 1$ compared to the case with $L=0$. This better performance, which is also observed in Fig. \ref{fig4}, is justified by the fact that, a better compensation process leads to a higher detection probability. Moreover, the results in Fig. \ref{fig6} indicate that, by using a more powerful compensation process the system performance becomes more robust to the degradations caused by the interruptions and reporting channel contaminations. For instance, we observe that the system performance degradation caused by increasing the energy efficiency level from 30\% to 50\% is smaller when a compensator with $L = 1$ is used compared to the case in which a compensator with $L = 0$ is considered. This is the advantage of using a more powerful compensation process at the FC. 

Since in the optimal linear fusion scheme  \cite{Taricco11, Quan10, Quan08} only the spatial correlations are used at the FC, the CROC curves designated by $L = 0$ in Fig. \ref{fig6} can be used to compare the proposed system performance with the performance of the optimal linear fusion. Specifically, it is clear in Fig. \ref{fig6} that, the optimal linear combining performance is closely achieved by the proposed system. However, this performance is obtained while the energy consumed for cooperative sensing is significantly reduced by the proposed architecture. The energy efficiency level obtained is even more significant when temporal correlations are also taken into account as demonstrated by CROC curves corresponding to $L = 1$. 

\begin{figure}[]
	\centering 
  \includegraphics[scale=0.30]{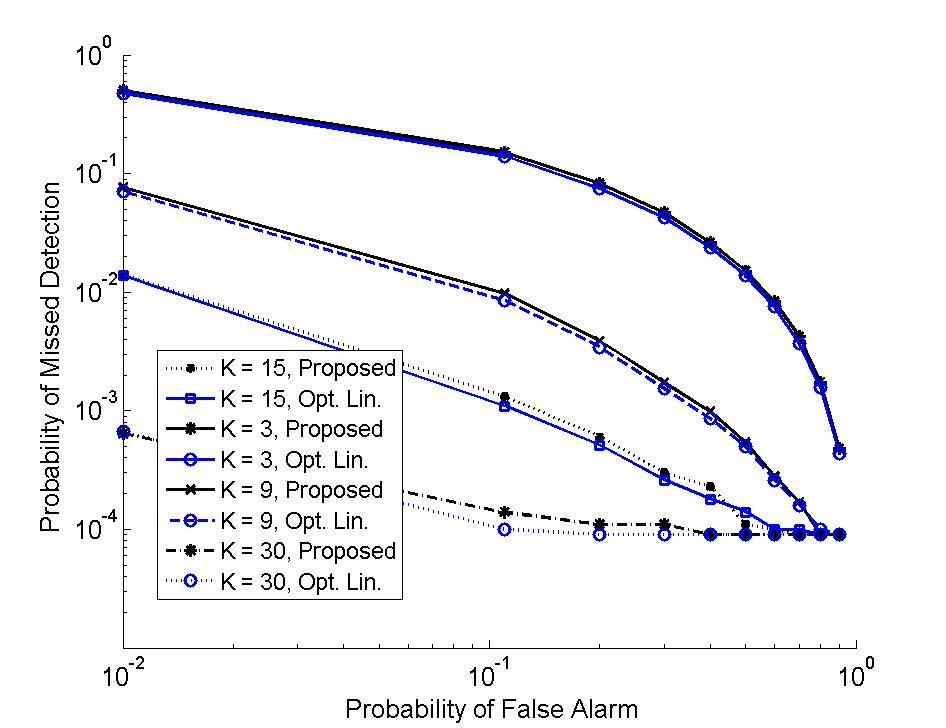} 
  \caption{Performance of the proposed SDP for the interrupted cooperation with 30\% energy efficiency compared with the performance of the optimal linear fusion scheme for $K = 3, 9, 15, 30$.  }
  \label{fig7}
\end{figure}

In Fig. \ref{fig7}, the results of converting the fractional program \eqref{SOP8} into a set of low-complexity SDPs is evaluated. We have considered three sets of nodes, operating at SNR levels as $\{12, 5, 7\}$ in dB. The efficiency level considered is 30\%. The CROC curves in this figure correspond to the cases where there are one, three, five, and ten nodes in each set of sensors. In this way, we investigate the system performance obtained by the proposed low-complexity suboptimal approach while increasing the number of sensing nodes. In addition, for each case the CROC curve corresponding to the optimal linear fusion is plotted for comparison. We can see in Fig. \ref{fig7} that the results obtained by the proposed relaxation approach are very close to the overall detection performance obtained by the optimal linear fusion scheme.

\begin{figure}[]
	\centering 
  \includegraphics[scale=0.30]{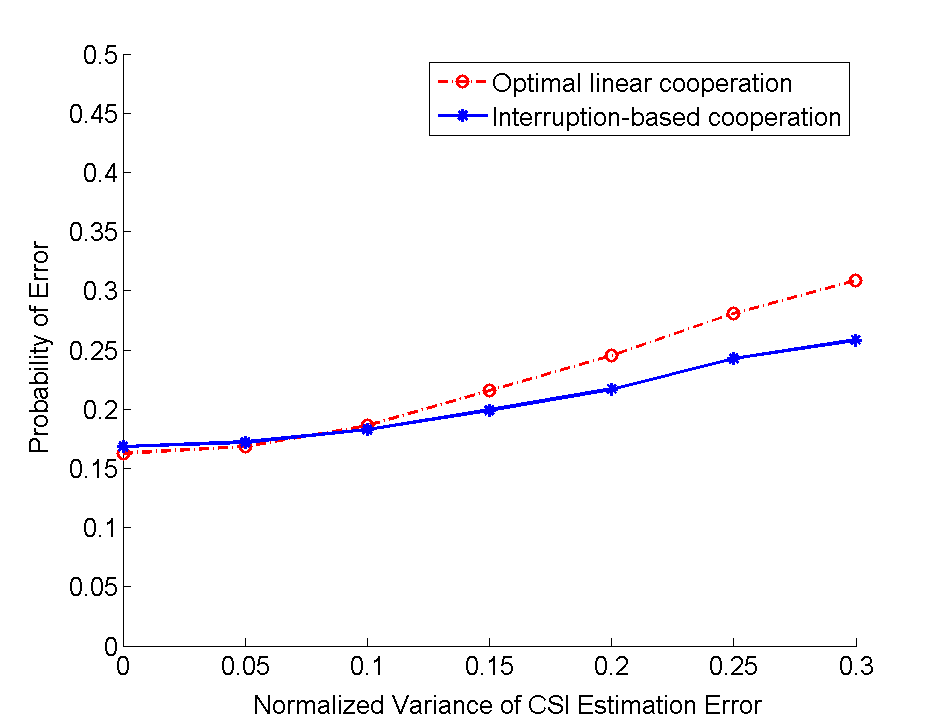} 
  \caption{Effect of imperfections in the available CSI on the probability of error in the proposed system and optimal linear fusion scheme. }
  \label{fig8}
\end{figure}

The effect of imperfect CSI on the overall detection performance is illustrated in Fig. \ref{fig8}. The proposed system performance is compared against the performance of the optimal linear combing scheme operating with the same erroneous CSI. The erroneous estimated first- and second-order channel statistics $\boldsymbol{\mu}_{{\mathbf{u}}| \mathcal{H}_h } $ and $\mathbf{C}_{{\mathbf{u}}| \mathcal{H}_h }$ (for $h = 0,1$) are realized as normally-distributed random variables centered at their corresponding actual values while different  estimation qualities are realized by changing the variances of these normal random variables. Three nodes operating at SNR levels $\{12, 5, 7\}$ in dB are considered in this simulation. These nodes generate independent local sensing results as in \cite{Quan10, Quan08}. Assuming independent sensing outcomes leads to diagonal covariance matrices, i.e., less number of parameters to be estimated, and consequently, faster Monte Carlo simulations. Since imperfections in the available CSI affects both the false alarm and detection probabilities, we compare the two cooperation methods by inspecting their probability of error defined as $P_e \triangleq \textup{Pr}\{\mathcal{H}_0\}P_f + \textup{Pr}\{\mathcal{H}_1\}(1-P_d)$. In Fig. \ref{fig8}, $P_e$ is depicted vs. the normalized variance of estimation error, which is the variance of each parameter to be estimated devided by its squared value. The results show that as expected, performance of both detectors are degraded when the CSI error increases. However, the proposed interrupted cooperation outperforms the optimal linear fusion scheme when operating with imperfect CSI and particularly when the error in CSI channel estimation is relatively high.

\begin{figure}[]
	\centering 
  \includegraphics[scale=0.30]{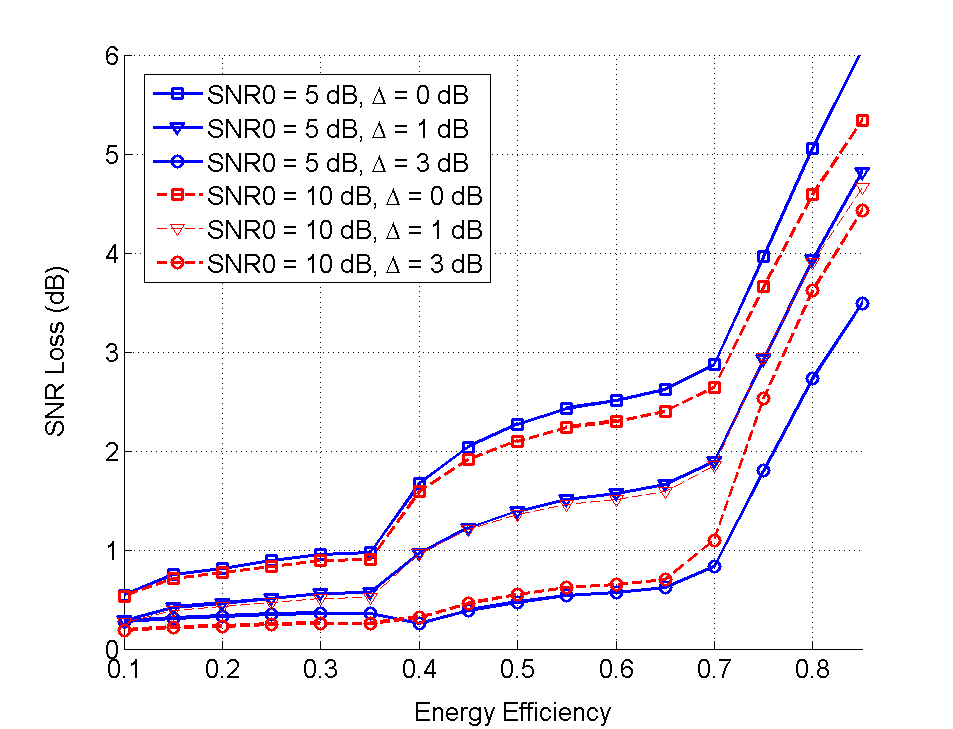} 
  \caption{SNR loss vs. energy efficiency for $P_{\textup{fa}} = 0.1$ obtained via the proposed cooperative sensing scheme with $K=3$ sensing nodes operating at two different average SNRs, i.e., $\textup{SNR}_0 = 5, 10$ in dB, with dispersion levels as $\Delta = 0,1,3$ in dB.}
  \label{fig9}
\end{figure}

The effect of SNR dispersion among the sensing nodes can be found in Fig. \ref{fig9}. In this figure, the  set of SNR levels experienced by the sensing nodes is $\{\textup{SNR}_0 + \Delta, \textup{SNR}_0, \textup{SNR}_0-\Delta\}$, where $\textup{SNR}_0$ denotes the average SNR of the sensing nodes and $\Delta$ represents the SNR dispersion.  Two cases for the average SNR are considered, each with three values for the dispersion. That is, $\textup{SNR}_0 = 5, 10$ in dB and $ \Delta=0,1,3$ in dB. It is clear that, a higher value for  $\Delta$ represents a greater difference in the conditions in which the sensing nodes work. In order to compare the proposed system performance with the performance of the optimal linear combining, we consider the loss in SNR defined as
\begin{eqnarray}\textup{SNR}~\textup{loss}~(\textup{dB})=\textup{SNR}_{\boldsymbol{\theta} }~(\textup{dB})-\textup{SNR}_{\mathbf{w}}~(\textup{dB})
\end{eqnarray}
where $\textup{SNR}_{\boldsymbol{\theta}}$ and $\textup{SNR}_{\mathbf{w}}$ denote the minimum SNR required at the SUs, incorporating the proposed method and optimal linear combining respectively, to meet a certain probability of false alarm, i.e., $P_{\textup{f}}=\alpha$, while having the same probability of missed detection. 

Fig. \ref{fig9} depicts the SNR loss vs. energy efficiency for $\alpha = 0.1$. As shown in this figure, by increasing the efficiency level, the SNR loss is increased. This behavior illustrates the discussed tradeoff between the sensing quality and the energy consumption. In addition, we see that the SNR loss is decreased as the SNR dispersion increases. Therefore, a better performance is achieved when there is a greater difference in the conditions experienced by the SUs, i.e., when there is a higher need to discriminate between different sensing outcomes. In all cases depicted in Fig. \ref{fig9} we observe that,  significant energy efficiency can be achieved while the detection performance is not heavily compromised. Moreover, the obtained energy efficiency is more significant when the dispersion is high. This is due to the fact that, when the SNR differences are high, the sensing outcomes from the nodes with low SNR do not considerably improve the overall detection, and therefore, the system suppresses their contribution by forcing them to go to the sleep mode more frequently as the efficiency level increases. Since the other nodes are relatively reliable, the overall detection performance is not affected significantly by this sleeping process. 

As another important observation, there are two breaking points on each curve in Fig. \ref{fig9}. By increasing the efficiency level up to the first breaking point, the proposed system suppresses the adverse effect of the hidden node on the overall sensing process. Up to this point, the degradations caused by the interruptions and reporting channel are compensated by the linear estimator at the FC. The first breaking point is nearly where we have established an appropriate level of discrimination between the reliable and unreliable nodes and any further power reduction degrades the system performance since it affects the reliable nodes as well.  Nevertheless, the compensator is still able to alleviate, to some extent, the degradations and consequently, the SNR loss does not grow as fast as it does when the second breaking point is passed. The second breaking point is nearly where the received reports quality is so low that the compensator is no longer able to reduce the performance degradation. Therefore, any further power reduction directly leads to significant loss of the detection quality. 
\section{Conclusion}\label{section Conclusion}
In this paper, a novel energy-efficient structure for spectrum sensing in CRNs has been proposed based on making random interruptions in the cooperation process among the CRs. The proposed system has been thoroughly modeled and analyzed, and an optimization problem has been developed in order to formulate a tradeoff taking into account the energy consumption at the local sensing and reporting processes jointly with the overall detection performance. Analytical solution of the optimization problem and the presented numerical results demonstrate that, significant levels of energy efficiency can be achieved by the proposed architecture. This energy efficiency is due to the fact that, unlike in existing cooperative sensing schemes, in the proposed design the discrimination between the reliable and unreliable nodes is obtained while no energy is wasted. Moreover, sensitivity of the overall detection to degradations in the local sensing and reporting processes is significantly reduced by the proposed architecture, leading to a more reliable cooperative spectrum sensing.

\section*{Appendix I\\ Second-Order Statistics of the Local Sensing Outcomes}\label{Appendix1}
In the following, we derive $\mathbf{C}_{\mathbf{u_L}}$ and $\boldsymbol{\mathfrak{R}}_{\mathbf{u_L}}$ in terms of the listening-channel and PU-signal statistics. $\mathbf{C}_{\mathbf{u}_L \mathbf{u}}$ can be  obtained similarly. We have  
\begin{eqnarray}\label{eq02201}
\left (\mathbf{C}_{\mathbf{u_L}}  \right )_{i,j} = \left (\mathbf{R}_{\mathbf{u_L}}  \right )_{i,j}-E\left [ u_k(m-l) \right ]E\left [ u_n(m-
r) \right ] 
\end{eqnarray}
where 
\begin{align}\label{eq022011} 
E&\left [ u_k(m-l) \right ] \\\nonumber 
&=  E \left [\sum_{i'=0}^{N-1} \left | h_ks(N(m-l)-i')+ \nu_k(N(m-l)-i')  \right |^2  \right ]\\\nonumber 
&= \sum_{i'=0}^{N-1}E\left [\left | h_ks(N(m-l)-i')+ \nu_k(N(m-l)-i')  \right |^2  \right ]\\\nonumber
& = N \left (E[g_k]\overline{|s|^2}+\sigma_{\nu}^2  \right ) 
\end{align}
and similarly, $E\left [ u_n(m-r) \right ] =N \left (E[g_n]\overline{|s|^2}+\sigma_{\nu}^2  \right )$, where $g_k \triangleq |h_k|^2$, $g_n \triangleq |h_n|^2$ are the channel power gains and $\overline{\left |s  \right |^2} \triangleq E[ s^2(m) ]$. Moreover, 
\begin{align}\label{eqR1551} 
&\left (\mathbf{R}_{\mathbf{u_L}}  \right )_{i,j}=E\left [ u_k(m-l)   u_n(m-
r) \right ] \\\nonumber
&= \sum_{i'=0}^{N-1}\sum_{j'=0}^{N-1} E\left [  \left | x_k(N(m-l)-i') \right |^2  \left | x_n(N(m-r)-j') \right |^2 \right ] \\\nonumber
&= \sum_{i'=0}^{N-1}\sum_{j'=0}^{N-1} 
\{ E\left [ g_kg_n\right ]R_{\left |s  \right |^2 } (N(l-r)+i'-j')
\\\nonumber 
&+ \sigma_{\nu}^2 \overline{\left |s  \right |^2}  \left ( E\left [ g_k\right ]  + E\left [ g_n\right ] \right )+ \sigma_{\nu}^4 \left ( 1+ \delta_{k=n}\delta_{N(l-r),i'-j'} \right )  \}
\end{align}
in which $R_{ |s |^2 } (l-r) \triangleq E[|s(m-l)|^2|s(m-r)|^2]$.  We have assumed, without loss of generality that, the signals are wide-sense stationary (WSS). In addition, the noise variances are assumed to be the same for all sensors i.e., $\sigma_{\nu_i} = \sigma_{\nu}$ for all $i$. Note that $\boldsymbol{\mathfrak{R}}_{\mathbf{u_L}}$ is a diagonal matrix whose main diagonal equals to the main diagonal of $\mathbf{R}_{\mathbf{u_L}}$.

\bibliographystyle{IEEEtran}
\bibliography{IEEEabrv,Bibliogeraphy_ICLAS}

\begin{thebibliography}{10}
\providecommand{\url}[1]{#1}
\csname url@samestyle\endcsname
\providecommand{\newblock}{\relax}
\providecommand{\bibinfo}[2]{#2}
\providecommand{\BIBentrySTDinterwordspacing}{\spaceskip=0pt\relax}
\providecommand{\BIBentryALTinterwordstretchfactor}{4}
\providecommand{\BIBentryALTinterwordspacing}{\spaceskip=\fontdimen2\font plus
\BIBentryALTinterwordstretchfactor\fontdimen3\font minus
  \fontdimen4\font\relax}
\providecommand{\BIBforeignlanguage}[2]{{%
\expandafter\ifx\csname l@#1\endcsname\relax
\typeout{** WARNING: IEEEtran.bst: No hyphenation pattern has been}%
\typeout{** loaded for the language `#1'. Using the pattern for}%
\typeout{** the default language instead.}%
\else
\language=\csname l@#1\endcsname
\fi
#2}}
\providecommand{\BIBdecl}{\relax}
\BIBdecl

\bibitem{mitola2000}
J.~Mitola, \emph{Cognitive Radio---An Integrated Agent Architecture for
  Software Defined Radio}.\hskip 1em plus 0.5em minus 0.4em\relax Royal
  Institute of Technology (KTH), 2000.

\bibitem{Akyildiz11}
I.~F. Akyildiz, B.~F. Lo, and R.~Balakrishnan, ``Cooperative spectrum sensing
  in cognitive radio networks: {A} survey,'' \emph{Elsevier Physical Commun.},
  vol.~4, no.~1, pp. 40--62, March 2011.

\bibitem{Paysarvi-Hoseini11}
P.~Paysarvi-Hoseini and N.~Beaulieu, ``Optimal wideband spectrum sensing
  framework for cognitive radio systems,'' \emph{Signal Processing, IEEE
  Transactions on}, vol.~59, no.~3, pp. 1170--1182, March 2011.

\bibitem{Abdi14}
Y.~Abdi and T.~Ristaniemi, ``Joint local quantization and linear cooperation in
  spectrum sensing for cognitive radio networks,'' \emph{Signal Processing,
  IEEE Transactions on}, vol.~62, no.~17, pp. 4349--4362, Sept 2014.

\bibitem{Chaudhari13}
S.~Chaudhari, J.~Lund\'{e}n, V.~Koivunen, and H.~V. Poor, ``{BEP} walls for
  cooperative sensing in cognitive radios using {K}-out-of-{N} fusion rules,''
  \emph{Elsevier Signal Process}, vol.~93, no.~7, pp. 1900--1908, July 2013.

\bibitem{Taricco11}
G.~Taricco, ``Optimization of linear cooperative spectrum sensing for cognitive
  radio networks,'' \emph{IEEE J. Sel. Topics Signal Process.}, vol.~5, no.~1,
  pp. 77--86, Feb. 2011.

\bibitem{Quan10}
Z.~Quan, W.~Ma, S.~Cui, and A.~H. Sayed, ``Optimal linear fusion for
  distributed detection via semidefinite programming,'' \emph{IEEE Trans.
  Signal Process.}, vol.~58, no.~4, pp. 2431--2436, 2010.

\bibitem{Quan08}
Z.~Quan, S.~Cui, and A.~H. Sayed, ``Optimal linear cooperation for spectrum
  sensing in cognitive radio networks,'' \emph{IEEE J. Sel. Topics Signal
  Process.}, vol.~2, no.~1, pp. 28--40, Feb. 2008.

\bibitem{Rago96}
C.~Rago, P.~Willett, and Y.~Bar-Shalom, ``Censoring sensors: a
  low-communication-rate scheme for distributed detection,'' \emph{Aerospace
  and Electronic Systems, IEEE Transactions on}, vol.~32, no.~2, pp. 554--568,
  April 1996.

\bibitem{Lunden09}
J.~Lund\'{e}n, V.~Koivunen, A.~Huttunen, and H.~V. Poor, ``Collaborative
  cyclostationary spectrum sensing for cognitive radio systems,'' \emph{IEEE
  Trans. Signal Process.}, vol.~57, no.~11, pp. 4182--4195, Nov. 2009.

\bibitem{Chen10}
Y.~Chen, ``Analytical performance of collaborative spectrum sensing using
  censored energy detection,'' \emph{Wireless Communications, IEEE Transactions
  on}, vol.~9, no.~12, pp. 3856--3865, December 2010.

\bibitem{Appad08}
S.~Appadwedula, V.~Veeravalli, and D.~Jones, ``Decentralized detection with
  censoring sensors,'' \emph{Signal Processing, IEEE Transactions on}, vol.~56,
  no.~4, pp. 1362--1373, April 2008.

\bibitem{Maleki13}
S.~Maleki, S.~P. Chepuri, and G.~Leus, ``Optimization of hard fusion based
  spectrum sensing for energy-constrained cognitive radio networks,''
  \emph{Physical Communication}, vol.~9, pp. 193 -- 198, 2013.

\bibitem{Maleki13-2}
S.~Maleki and G.~Leus, ``Censored truncated sequential spectrum sensing for
  cognitive radio networks,'' \emph{Selected Areas in Communications, IEEE
  Journal on}, vol.~31, no.~3, pp. 364--378, March 2013.

\bibitem{Maleki11}
S.~Maleki, A.~Pandharipande, and G.~Leus, ``Energy-efficient distributed
  spectrum sensing for cognitive sensor networks,'' \emph{Sensors Journal,
  IEEE}, vol.~11, no.~3, pp. 565--573, March 2011.

\bibitem{Maleki15}
S.~Maleki, G.~Leus, S.~Chatzinotas, and B.~Ottersten, ``To and or to or: On
  energy-efficient distributed spectrum sensing with combined censoring and
  sleeping,'' \emph{Wireless Communications, IEEE Transactions on}, vol.~PP,
  no.~99, pp. 1--1, 2015.

\bibitem{Xia10}
W.~Xia, W.~Yuan, W.~Cheng, W.~Liu, S.~Wang, and J.~Xu, ``Optimization of
  cooperative spectrum sensing in ad-hoc cognitive radio networks,'' in
  \emph{Global Telecommunications Conference (GLOBECOM 2010), 2010 IEEE}, Dec
  2010, pp. 1--5.

\bibitem{Pham10}
H.~N. Pham, Y.~Zhang, P.~Engelstad, T.~Skeie, and F.~Eliassen, ``Energy
  minimization approach for optimal cooperative spectrum sensing in
  sensor-aided cognitive radio networks,'' in \emph{Wireless Internet
  Conference (WICON), 2010 The 5th Annual ICST}, March 2010, pp. 1--9.

\bibitem{Najimi13}
M.~Najimi, A.~Ebrahimzadeh, S.~Andargoli, and A.~Fallahi, ``A novel sensing
  nodes and decision node selection method for energy efficiency of cooperative
  spectrum sensing in cognitive sensor networks,'' \emph{Sensors Journal,
  IEEE}, vol.~13, no.~5, pp. 1610--1621, May 2013.

\bibitem{Monemian14}
M.~Monemian and M.~Mahdavi, ``Analysis of a new energy-based sensor selection
  method for cooperative spectrum sensing in cognitive radio networks,''
  \emph{Sensors Journal, IEEE}, vol.~14, no.~9, pp. 3021--3032, Sept 2014.

\bibitem{Xing15}
X.~Liu, B.~Evans, and K.~Moessner, ``Energy-efficient sensor scheduling
  algorithm in cognitive radio networks employing heterogeneous sensors,''
  \emph{Vehicular Technology, IEEE Transactions on}, vol.~64, no.~3, pp.
  1243--1249, March 2015.

\bibitem{Cardei05}
M.~Cardei, M.~Thai, Y.~Li, and W.~Wu, ``Energy-efficient target coverage in
  wireless sensor networks,'' in \emph{INFOCOM 2005. 24th Annual Joint
  Conference of the IEEE Computer and Communications Societies. Proceedings
  IEEE}, vol.~3, March 2005, pp. 1976--1984 vol. 3.

\bibitem{Jiming10}
J.~Chen, J.~Li, S.~He, Y.~Sun, and H.-H. Chen, ``Energy-efficient coverage
  based on probabilistic sensing model in wireless sensor networks,''
  \emph{Communications Letters, IEEE}, vol.~14, no.~9, pp. 833--835, September
  2010.

\bibitem{Lee11}
J.-W. Lee, B.-S. Choi, and J.-J. Lee, ``Energy-efficient coverage of wireless
  sensor networks using ant colony optimization with three types of
  pheromones,'' \emph{Industrial Informatics, IEEE Transactions on}, vol.~7,
  no.~3, pp. 419--427, Aug 2011.

\bibitem{Gil11}
J.-M. Gil and Y.-H. Han, ``A target coverage scheduling scheme based on genetic
  algorithms in directional sensor networks,'' \emph{Sensors}, vol.~11, no.~2,
  p. 1888, 2011.

\bibitem{Aalo92}
V.~Aalo and R.~Viswanathan, ``Asymptotic performance of a distributed detection
  system in correlated gaussian noise,'' \emph{Signal Processing, IEEE
  Transactions on}, vol.~40, no.~1, pp. 211--213, Jan 1992.

\bibitem{Ghasemi05}
A.~Ghasemi and E.~Sousa, ``Collaborative spectrum sensing for opportunistic
  access in fading environments,'' in \emph{New Frontiers in Dynamic Spectrum
  Access Networks, 2005. DySPAN 2005. 2005 First IEEE International Symposium
  on}, Nov 2005, pp. 131--136.

\bibitem{Visotsky05}
E.~Visotsky, S.~Kuffner, and R.~Peterson, ``On collaborative detection of tv
  transmissions in support of dynamic spectrum sharing,'' in \emph{Proc. Int.
  Conf. Dynam. Spectrum Access Net.}, Baltimore, MD, Nov. 8-11 2005, pp.
  2947--2951.

\bibitem{Poor}
H.~V. Poor, \emph{An introduction to signal detection and estimation}.\hskip
  1em plus 0.5em minus 0.4em\relax Berlin Heidelberg: Springer-Verlag, 1994.

\bibitem{Chen05}
B.~Chen and P.~K.Willett, ``On the optimality of the likelihood-ratio test for
  local sensor decision rules in the presence of nonideal channels,''
  \emph{IEEE Trans. Inform. Theory}, vol.~51, pp. 693--699, Feb. 2005.

\bibitem{Chaudhari12}
S.~Chaudhari, J.~Lunden, V.~Koivunen, and H.~V. Poor, ``Cooperative sensing
  with imperfect reporting channels: {H}ard decisions or soft decisions?''
  \emph{IEEE Trans. Signal Process.}, vol.~60, no.~1, pp. 18--28, Jan. 2012.

\bibitem{Chaudhari11}
S.~Chaudhari and V.~K. J.~Lunden, ``{BEP} walls for collaborative spectrum
  sensing,'' in \emph{Proc. IEEE International Conf. on Acoustics, Speech and
  Signal Processing (ICASSP)}, May 22-27 2011, pp. 2984--2987.

\bibitem{Chaudhari09}
S.~Chaudhari, , and V.~Koivunen, ``Effect of quantization and channel errors on
  collaborative spectrum sensing,'' in \emph{Proc. The 43rd Asilomar Conference
  on Signals, Systems, and Computers}, Nov. 2009, pp. 528--533.

\bibitem{Picinbono95}
B.~Picinbono, ``On deflection as a performance criterion in detection,''
  \emph{Aerospace and Electronic Systems, IEEE Transactions on}, vol.~31,
  no.~3, pp. 1072--1081, Jul 1995.

\bibitem{Derakhshani11}
M.~Derakhshani, T.~Le-Ngoc, and M.~Nasiri-Kenari, ``Efficient cooperative
  cyclostationary spectrum sensing in cognitive radios at low {SNR} regimes,''
  \emph{IEEE Trans. Wireless Commun.}, vol.~10, no.~11, pp. 3754--3764, Nov.
  2011.

\bibitem{Quan09}
Z.~Quan, S.~Cui, A.~H. Sayed, and H.~V. Poor, ``Optimal multiband joint
  detection for spectrum sensing in cognitive radio networks,'' \emph{IEEE
  Trans. Signal Process.}, vol.~57, no.~3, pp. 1128--1140, March 2009.

\bibitem{AbdiPIMRC14}
Y.~Abdi and T.~Ristaniemi, ``Extension of deflection coefficient for linear
  fusion of quantized reports in cooperative sensing,,'' in \emph{Proc. IEEE
  PIMRC}, Washington DC, USA, Sep. 2-5 2014.

\bibitem{Appad05}
S.~Appadwedula, V.~Veeravalli, and D.~Jones, ``Energy-efficient detection in
  sensor networks,'' \emph{Selected Areas in Communications, IEEE Journal on},
  vol.~23, no.~4, pp. 693--702, April 2005.

\bibitem{Chunhua07}
C.~Sun, W.~Zhang, and K.~Letaief, ``Cooperative spectrum sensing for cognitive
  radios under bandwidth constraints,'' in \emph{Wireless Communications and
  Networking Conference, 2007.WCNC 2007. IEEE}, March 2007, pp. 1--5.

\bibitem{Cordeiro06}
C.~Cordeiro, K.~Challapali, D.~Birru, and S.~Shankar~N, ``{IEEE} 802.22: An
  introduction to the first wireless standard based on cognitive radios,''
  \emph{Journal of Communications}, vol.~1, no.~1, pp. 38--47, 2006.

\bibitem{Defourny11}
B.~Defourny, D.~Ernst, and L.~Wehenkel, ``Multistage stochastic programming: A
  scenario tree based approach,'' \emph{Decision Theory Models for Applications
  in Artificial Intelligence: Concepts and Solutions: Concepts and Solutions},
  p.~97, 2011.

\bibitem{Benson06}
H.~P. Benson, ``Maximizing the ratio of two convex functions over a convex
  set,'' \emph{Naval Research Logistics}, vol.~53, no.~4, pp. 309--317, June
  2006.

\bibitem{Bao11}
X.~Bao, N.~Sahinidis, and M.~Tawarmalani,
  ``\BIBforeignlanguage{English}{Semidefinite relaxations for quadratically
  constrained quadratic programming: A review and comparisons},''
  \emph{\BIBforeignlanguage{English}{Mathematical Programming}}, vol. 129,
  no.~1, pp. 129--157, 2011.

\bibitem{shor90}
N.~Shor, ``Dual quadratic estimates in polynomial and boolean programming,''
  \emph{Annals of Operations Research}, vol.~25, no.~1, pp. 163--168, 1990.

\bibitem{anstreicher09}
K.~M. Anstreicher, ``Semidefinite programming versus the
  reformulation-linearization technique for nonconvex quadratically constrained
  quadratic programming,'' \emph{Journal of Global Optimization}, vol.~43, no.
  2-3, pp. 471--484, 2009.

\bibitem{Bienstock14}
D.~Bienstock and A.~Michalka, ``Polynomial solvability of variants of the
  trust-region subproblem,'' in \emph{Proceedings of the Twenty-Fifth Annual
  ACM-SIAM Symposium on Discrete Algorithms}, ser. SODA '14.\hskip 1em plus
  0.5em minus 0.4em\relax SIAM, 2014, pp. 380--390.

\end{thebibliography}

\end{document}